\documentclass[12pt]{report}

\usepackage{amsmath}
\usepackage{amsfonts}
\usepackage{mathrsfs}
\usepackage{mathtools}

\begin{document}


\thispagestyle{empty}

\vspace*{0.5cm}

\begin{center}
{\Large\bf Linearized Analysis of Rastall Gravity}
\end{center}

\vspace{3in}

\begin{center}
Miss Yuwadee Tongkong
\end{center}

\vspace{3in}

\begin{center}
A report submitted to the Department of Physics of Chulalongkorn University \\
in partial fulfillment of the requirements for the degree of \\
Bachelor of Science in Physics \\
Academic Year 2024
\end{center}

\clearpage


\pagenumbering{roman}
\setcounter{page}{1}

\noindent Project Title: ~~~~~~~Linearized Analysis of Rastall Gravity \\
\noindent Name: ~~~~~~~~~~~~~~~~Miss Yuwadee Tongkong \\
\noindent Project Advisor: ~~~Dr. Rujikorn Dhanawittayapol \\
\noindent Department: ~~~~~~~~Physics \\
\noindent Academic Year: ~~~~2024

\vspace{0.5cm}

\noindent\hrulefill

\vspace{0.5cm}

\begin{center}
{\bf ABSTRACT}
\end{center}

\vspace{0.3cm}

~~~~~Rastall gravity is a generalization of the Einstein 
gravity in which the matter is not conserved in the presence 
of a non-constant spacetime curvature. In this report, we 
analyze Rastall gravity using the linearized formalism. 
The linearized metric for a localized matter without pressure 
and shear stress is obtained, and a spherically symmetric 
metric for a non-rotating mass is derived. A phenomenological 
consequence, known as the gravitoelectromagnetism, is 
subsequently discussed in detail, where it is shown that 
a free-falling observer will see the motion of a nearby 
free particle as being subject to a velocity-independent 
gravitoelectric force and a velocity-dependent gravitomagnetic 
force. An explicit calculation of the gravitoelectric 
and gravitomagnetic fields as seen by an observer moving 
in a circular orbit in a spherically symmetric gravitational 
field is presented, and it is shown that the resulting 
gravitomagnetic field is perpendicular to the observer's 
moving direction.

\clearpage

\tableofcontents
\clearpage


\pagenumbering{arabic}
\setcounter{page}{1}


\chapter{Introduction}

General Relativity is a theory that describes 
gravity in terms of the spacetime curvature as 
related to the distribution of matter by the 
so-called Einstein's equation,
\begin{equation}
R_{\mu\nu} - \frac{1}{2}g_{\mu\nu}R = 
\frac{8\pi G}{c^4}T_{\mu\nu},                \label{1.1}
\end{equation}
where $R_{\mu\nu}$ is the Ricci tensor, $R$ 
is the curvature scalar, and $T_{\mu\nu}$ is 
the energy-momentum-stress tensor. The left-hand 
side describes the spacetime curvature, while 
the right-hand side describes the distribution 
of matter which causes the spacetime geometry 
to become curved. That the left-hand side of 
this equation takes the above form is due to 
the fact that it satisfies the identity 
\begin{equation}
\nabla^\mu (R_{\mu\nu} - 
\frac{1}{2}g_{\mu\nu}R) = 0,                \label{1.2}
\end{equation}
which is exactly the same as the desired 
property of the energy-momentum-stress tensor, 
\begin{equation}
\nabla^\mu T_{\mu\nu} = 0,                  \label{1.3}
\end{equation}
which says that the energy and momentum of matter 
must be conserved in the sense that the change 
of energy and momentum inside a region in space 
is equal to the net flow of energy and momentum 
into the region. In other words, the property 
$\nabla^\mu T_{\mu\nu} = 0$ simply says that 
the energy and momentum cannot pop up from 
nowhere. 

In 1972, P. Rastall proposed a modified version 
of General Relativity \cite{Rastall}, in which 
the existence of non-constant curvature scalar 
alters the law of conservation of energy and 
momentum in the manner that 
\begin{equation}
\nabla^\mu T_{\mu\nu} \sim \nabla_\nu R,    \label{1.4}
\end{equation}
thereby changing the Einstein's equation to 
\begin{equation}
(R_{\mu\nu} - \frac{1}{2}g_{\mu\nu}R) 
+ \lambda g_{\mu\nu}R = 
\frac{8\pi G}{c^4}T_{\mu\nu},               \label{1.5}
\end{equation}
with $\lambda$ being a constant. This leads to 
the modified conservation law, 
\begin{equation}
\frac{8\pi G}{c^4}\nabla^\mu T_{\mu\nu} = 
\lambda\nabla_\nu R,                       \label{1.6}
\end{equation}
which says that the energy-momentum-stress 
tensor is no longer conserved in the region 
in which $\nabla_\mu R\ne 0$. In other words, 
if Rastall's proposal is correct at least 
approximately, Eq. (\ref{1.6}) will result 
in the ``popping-up of matter from nowhere" 
or the ``spontaneous disappearance of matter" 
due to the non-constant spacetime curvature. 
Due to the past success of General Relativity, 
the $\lambda$ term in Eq. (\ref{1.5}) cannot 
be too large, and so it is interesting to 
estimate the maximum order of magnitude of 
the constant $\lambda$. To do this, let us 
consider the gravitational field nearby the 
Earth surface. Assuming that the spacetime 
structure in the vicinity of the Earth can be 
approximately described by the Schwarzschild 
metric, it is well known that the curvature 
scalar nearby the Earth surface is of the order 
\begin{equation}
R \sim \frac{GM_E}{R_E^3c^2},               \label{1.7}
\end{equation}
where $M_E$ and $R_E$ are, respectively, the 
Earth mass and the Earth radius. From Eq. 
(\ref{1.7}), it is clear that
\begin{equation}
\nabla_i R \sim \frac{GM_E}{R_E^4c^2}       \label{1.8}
\end{equation}
nearby the Earth surface. Thus, using Eqs. 
(\ref{1.6}) and (\ref{1.8}), we obtain
\begin{eqnarray}
\nabla_\mu T^{\mu\nu} &\sim & \lambda\left(
\frac{M_Ec^2}{8\pi R_E^4}\right)  \nonumber \\
&\approx & \left(10^{13}\right)\lambda ,    \label{1.9}
\end{eqnarray}
where we have used $M_E \approx 6 \times 10^{24}$ 
kg and $R_E \approx 6.4 \times 10^6$ m. 
This implies that the order of magnitude of 
$\lambda$ should be very much smaller than 
$10^{-13}$, otherwise all the physical theories 
that have been tested in the laboratories on 
Earth will be destroyed. 

In 1982, Lindblom and Hiscock \cite{Lindblom} 
obtained the upper bound of $\lambda$ to be 
less than $10^{-15}$. Their argument goes as 
follows. From Eq. (\ref{1.5}), one can obtain 
the trace of the energy-momentum-stress tensor 
in terms of the curvature scalar, and rewrite 
Eq. (\ref{1.5}) as
\begin{equation}
R_{\mu\nu} - \frac{1}{2}g_{\mu\nu}R = 
\frac{8\pi G}{c^4}S_{\mu\nu},               \label{1.10}
\end{equation}
where
\begin{equation}
S_{\mu\nu} \equiv T_{\mu\nu} + \frac{\lambda}
{1-4\lambda}g_{\mu\nu}T,                    \label{1.11}
\end{equation}
with $T \equiv g^{\mu\nu}T_{\mu\nu}$ being 
the trace of $T_{\mu\nu}$. Thus, $S_{\mu\nu}$ 
is conserved,
\begin{equation}
\nabla_\mu S^{\mu\nu} = 0.                  \label{1.12}
\end{equation}
Choosing the matter to be a perfect fluid with 
\begin{equation}
T_{\mu\nu} = (\varepsilon + P)u_\mu u_\nu 
- P g_{\mu\nu},                             \label{1.13}
\end{equation}
where $\varepsilon$ and $P$ are, respectively, 
the fluid energy density and the fluid pressure, 
Eq. (\ref{1.12}) leads to 
\begin{equation}
\frac{\partial}{\partial t}[(1-3\lambda)\varepsilon 
-3\lambda P] + (1-4\lambda)\varepsilon\nabla_i v^i 
= 0                                         \label{1.14}
\end{equation}
and 
\begin{equation}
(1-4\lambda)\varepsilon\left(
\frac{\partial v^i}{\partial t} + v^j\nabla_jv^i\right)
= -c^2\nabla^i[(1-\lambda)P - \lambda\varepsilon] 
                                            \label{1.15}
\end{equation}
in the non-relativistic limit, where $v^i$ is 
the fluid velocity. It is easy to see that 
Eqs. (\ref{1.14}) and (\ref{1.15}) are exactly 
of the same form as the fluid-dynamics equations 
describing the sound wave (see pages 251--252 
of \cite{Landau}), 
\begin{equation}
\frac{\partial(\delta\rho)}{\partial t} + 
\rho_0\vec{\nabla}\!\cdot\!\vec{v} = 0,     \label{1.16}
\end{equation}
\begin{equation}
\rho_0\frac{\partial\vec{v}}{\partial t} + 
\rho_0(\vec{v}\!\cdot\!\vec{\nabla})\vec{v} = 
-\vec{\nabla} p,                            \label{1.17}
\end{equation}
where $\rho_0$ is the equilibrium fluid density, 
$\delta\rho$ is the density fluctuation, and 
$p$ is the pressure. Since the speed of the 
sound wave can be obtained from 
\begin{equation}
v_s^2 = \frac{\partial p}{\partial(\delta\rho)}, 
                                            \label{1.18}
\end{equation}
the analogy between Eqs. (\ref{1.14})--(\ref{1.15}) 
and Eqs. (\ref{1.16})--(\ref{1.17}) leads to 
the expression for the speed of sound wave in 
Rastall gravity as 
\begin{equation}
v_s^2 = c^2\left( \frac{(1-\lambda)
\frac{\partial P}{\partial\varepsilon} - 
\lambda}{(1-3\lambda) - 3\lambda
\frac{\partial P}{\partial\varepsilon}} \right). 
                                           \label{1.19}
\end{equation}
When $\lambda$ is much less than the order 
of $10^{-13}$, Eq. (\ref{1.19}) reduces to 
\begin{equation}
v_s^2 = \frac{\partial P}{\partial(\delta\rho)} 
- \lambda c^2,                             \label{1.20}
\end{equation}
where we have used $\varepsilon = (\rho_0 + 
\delta\rho)c^2$. Thus, the speed of sound wave 
in Rastall gravity differs from the classical 
result by $\lambda c^2$. By setting $\lambda 
c^2$ to be equal to the experimental error in 
\cite{Itterbeek}, Lindblom and Hiscock were 
able to obtain the upper bound of $\lambda$ 
to be $|\lambda| < 10^{-15}$. 

It is interesting to ask if there is an 
alternative way to obtain an upper bound 
of $\lambda$ from experiments done in the 
vicinity of the Earth surface. As the gravity 
is weak nearby the Earth surface, it is 
reasonable to confine ourselves to the 
linearized Rastall gravity, in which one 
takes the spacetime metric to be a flat 
metric plus a small perturbation and keeps 
all the things up to the first order in 
perturbation \cite{Schutz}. This is what we 
will do in this report. As for the determination 
of the upper bound of $\lambda$, we will 
concentrate on the phenomenon of 
gravitoelectromagnetism in Rastall gravity, 
and investigate the effect of $\lambda$ on 
the measurable quantities to see if there 
is any way to obtain the upper bound of $\lambda$. 

The organization of this report is as follows. 
In Chapter 2, we will apply the linearized 
formalism of General Relativity to Rastall gravity 
to obtain the linearized equation for the metric 
perturbation. We next obtain the general solution 
to the linearized equation, and then evaluate 
it in the case of the matter without pressure 
and shear stress. The phenomenological consequence 
of the linearized Rastall gravity is subsequently 
discussed in Chapter 3, in which we begin with 
a review of the construction of the Fermi normal 
coordinate system, and then go on to use it to 
obtain the equation of motion of a free particle 
as seen by a free-falling observer in terms of 
the gravitoelectric and gravitomagnetic forces. 
Chapter 3 will end with the computation of the 
gravitoelectric and gravitomagnetic fields 
measured by an observer moving along a circular 
orbit and the discussion of the possibility of 
estimating the upper bound of $\lambda$. This 
report will end with Chapter 4, in which the 
conclusions are made.



\chapter{Linearized Rastall Gravity}

In this chapter, we will consider Rastall gravity 
\cite{Rastall} in the situation in which the gravitational 
field is weak, and so the spacetime curvature is small. 
We will present the general formalism in the first 
section, and then go on to consider a specific case 
of matter without pressure and shear stress. 


\section{General Formalism}

In the presence of a weak gravitation field, the 
spacetime must have a small curvature. This implies 
that there must exist a coordinate system in spacetime 
in which the metric deviates from the Minkowski metric 
by a small amount. In such a coordinate system, the 
metric tensor takes the form \cite{Schutz} 
\begin{equation}
g_{\mu\nu} = \eta_{\mu\nu} + h_{\mu\nu},        \label{2.1}
\end{equation}
where $\eta_{\mu\nu}$ is the Minkowski metric, which 
is chosen to be mostly minus ($\eta_{\mu\nu} = 
\mbox{diag}(1,-1,-1,-1)$), and $h_{\mu\nu}$ is a 
small metric perturbation, $|h_{\mu\nu}|\ll 1$. 
It is easy to verify that the corresponding inverse 
metric tensor is
\begin{equation}
g^{\mu\nu} = \eta^{\mu\nu} - \eta^{\mu\rho}
\eta^{\nu\sigma}h_{\rho\sigma},                \label{2.2}
\end{equation}
valid up to the linear order in the metric perturbation. 
Using Eqs. (\ref{2.1})--(\ref{2.2}), we can calculate 
the Christoffel connection, 
\begin{eqnarray}
\Gamma^\rho_{\mu\nu} &=& \frac{1}{2}g^{\rho\sigma}
(\partial_\mu g_{\nu\sigma} + \partial_\nu g_{\mu\sigma} 
- \partial_\sigma g_{\mu\nu})  \nonumber \\
&=& \frac{1}{2}\eta^{\rho\sigma}
(\partial_\mu h_{\nu\sigma} + \partial_\nu h_{\mu\sigma} 
- \partial_\sigma h_{\mu\nu}),                \label{2.3}
\end{eqnarray}
the Riemann curvature tensor,
\begin{eqnarray}
R^\mu{}_{\nu\rho\sigma} &=& \partial_\rho\Gamma^\mu_{\sigma\nu}
- \partial_\sigma\Gamma^\mu_{\rho\nu} + 
\Gamma^\mu_{\rho\tau}\Gamma^\tau_{\sigma\nu} - 
\Gamma^\mu_{\sigma\tau}\Gamma^\tau_{\rho\nu}  \nonumber \\
&=& \frac{1}{2}\eta^{\mu\tau}(
\partial_\rho\partial_\nu h_{\tau\sigma} + 
\partial_\tau\partial_\sigma h_{\rho\nu} - 
\partial_\sigma\partial_\nu h_{\tau\rho} - 
\partial_\tau\partial_\rho h_{\sigma\nu}),    \label{2.4}
\end{eqnarray}
the Ricci tensor,
\begin{eqnarray}
R_{\mu\nu} &=& R^\rho{}_{\mu\rho\nu}  \nonumber \\
&=& \frac{1}{2}(\partial^\rho\partial_\mu h_{\nu\rho} 
+ \partial^\rho\partial_\nu h_{\mu\rho} 
- \partial_\mu\partial_\nu h - \Box h_{\mu\nu}), 
                                             \label{2.5}
\end{eqnarray}
where $h \equiv \eta^{\mu\nu}h_{\mu\nu}$ and $\Box\equiv
\eta^{\mu\nu}\partial_\mu\partial_\nu$, and the scalar 
curvature,
\begin{eqnarray}
R &=& g^{\mu\nu}R_{\mu\nu}  \nonumber \\
&=& \partial^\mu\partial^\nu h_{\mu\nu} - \Box h, 
                                             \label{2.6}
\end{eqnarray}
valid up to the linear order in $h_{\mu\nu}$. 

Substitute the above result into the Rastall-gravity 
equation,
\begin{equation}
R_{\mu\nu} - \frac{1}{2}g_{\mu\nu}R + \lambda g_{\mu\nu}R 
= \frac{8\pi G}{c^4}\,T_{\mu\nu},            \label{2.7}
\end{equation}
the left-hand side of this equation takes the linearized 
form,
\begin{eqnarray}
R_{\mu\nu} - \frac{\tilde{\lambda}}{2}g_{\mu\nu}R &=& 
\frac{1}{2}(\partial^\rho\partial_\mu h_{\nu\rho} 
+ \partial^\rho\partial_\nu h_{\mu\rho} 
- \partial_\mu\partial_\nu h - \Box h_{\mu\nu}) \nonumber \\
&& - \frac{\tilde{\lambda}}{2}\eta_{\mu\nu}(
\partial^\rho\partial^\sigma h_{\rho\sigma} - \Box h), 
                                             \label{2.8}
\end{eqnarray}
where $\tilde{\lambda}\equiv (1-2\lambda)$. This looks 
a little bit messy, so one might ask if there is any 
way to simplify this result. The answer to this question 
lies in the invariance of the physics under the general 
coordinate transformation. Since we have our freedom 
to choose the coordinate system to work with without 
changing the physics, we would like to find a coordinate 
system in which the metric tensor still takes the 
form $(\ref{2.1})$ but with Eq. (\ref{2.8}) taking a 
simpler and manageable form. Indeed, let us consider 
an infinitesimal coordinate transformation, 
\begin{equation}
x^{\mu\prime} = x^\mu + \epsilon\xi^\mu ,         \label{2.9}
\end{equation}
where $\xi^\mu$ is some vector and $\epsilon$ is an 
infinitesimal parameter. The metric tensor changes 
according to the rule 
\begin{equation}
g^\prime_{\mu\nu}(x^\prime) = 
\frac{\partial x^\rho}{\partial x^{\mu\prime}}
\frac{\partial x^\sigma}{\partial x^{\nu\prime}}\,
g_{\rho\sigma}(x),                              \label{2.10}
\end{equation}
which results in the change of the metric perturbation 
to 
\begin{equation}
h^\prime_{\mu\nu} = h_{\mu\nu} - 
\epsilon\partial_\mu\xi_\nu - \epsilon
\partial_\nu\xi_\mu,                            \label{2.11}
\end{equation}
to the first order in $\epsilon$ and $h_{\mu\nu}$. 
The corresponding change in $h = \eta^{\mu\nu}
h_{\mu\nu}$ is 
\begin{equation}
h^\prime = h - 2\epsilon\partial_\mu\xi^\mu .   \label{2.12}
\end{equation}
Using 
\begin{eqnarray}
\partial^\prime_\mu &=& \left(\frac{\partial x^\nu}
{\partial x^{\mu\prime}}\right)\partial_\nu  \nonumber \\
&=& \partial_\mu - \epsilon 
\left(\partial_\mu\xi^\nu\right)\partial_\nu , 
                                                \label{2.13}
\end{eqnarray}
we calculate
\begin{equation}
\partial^\prime_\mu ( h^{\mu\nu\prime} - \frac{1}{2}
\eta^{\mu\nu}h^\prime ) = 
\partial_\mu ( h^{\mu\nu} - \frac{1}{2}
\eta^{\mu\nu}h ) - \epsilon\Box\xi^\nu          \label{2.14}
\end{equation}
to the first order in $\epsilon$ and $h_{\mu\nu}$, 
where we have used Eqs. (\ref{2.11})--(\ref{2.12}). 
Suppose we originally use the coordinate system 
$x^\mu$ in which $\partial_\mu (h^{\mu\nu}-\frac{1}{2}
\eta^{\mu\nu}h) \ne 0$, Eq. (\ref{2.14}) tells us 
that we can always change the coordinate system 
to $x^{\mu\prime}$ such that $\partial^\prime_\mu 
(h^{\mu\nu\prime}-\frac{1}{2}\eta^{\mu\nu}h^\prime) 
= 0$ by using $\xi^\mu$ in Eq. (\ref{2.9}) that 
satisfies 
\begin{equation}
\epsilon\Box\xi^\nu = \partial_\mu ( h^{\mu\nu} - 
\frac{1}{2}\eta^{\mu\nu}h),                  \label{2.15}
\end{equation}
which can always be solved once the explicit form 
of $\partial_\mu (h^{\mu\nu}-\frac{1}{2}\eta^{\mu\nu}h)$ 
is known. This enables us to impose the condition 
\begin{equation}
\partial_\mu (h^{\mu\nu}-\frac{1}{2}\eta^{\mu\nu}h) 
= 0                                          \label{2.16}
\end{equation}
on $h_{\mu\nu}$ (which can always be achieved by 
choosing a suitable coordinate system), and we 
will assume this condition from now on. With the 
condition (\ref{2.16}), Eq. (\ref{2.8}) becomes
\begin{equation}
R_{\mu\nu} - \frac{\tilde{\lambda}}{2}g_{\mu\nu}R 
= -\frac{1}{2}\Box\tilde{h}_{\mu\nu},        \label{2.17}
\end{equation}
where we have defined 
\begin{equation}
\tilde{h}_{\mu\nu} \equiv h_{\mu\nu} - 
\frac{\tilde{\lambda}}{2}\eta_{\mu\nu}h.     \label{2.18}
\end{equation}
Using the above result in Eq. (\ref{2.7}), we obtain 
the linearized equation of Rastall gravity,
\begin{equation}
\Box\tilde{h}_{\mu\nu} = -\frac{16\pi G}{c^4}T_{\mu\nu}, 
                                             \label{2.19}
\end{equation}
which takes the form of the wave equation with the 
energy-momentum-stress tensor as a source term. The 
solution to this equation is a sum of two parts. 
The first part is the complementary solution, 
$(\tilde{h}_c)_{\mu\nu}$, satisfying the homogeneous 
wave equation, $\Box (\tilde{h}_c)_{\mu\nu} = 0$. 
It is easy to see that $(\tilde{h}_c)_{\mu\nu}$ 
is a plane wave, 
\begin{equation}
(\tilde{h}_c)_{\mu\nu}(x) = \varepsilon_{\mu\nu}\, 
e^{ik \cdot x} + \varepsilon^\ast_{\mu\nu}
e^{-ik \cdot x},                            \label{2.20}
\end{equation}
where $\varepsilon_{\mu\nu}$ is a constant, and 
$k\!\cdot\!x = \eta_{\mu\nu}k^\mu x^\nu$ with 
$k^\mu$ being a wave vector satisfying $k^2 = 
\eta_{\mu\nu}k^\mu k^\nu = 0$. Since this solution 
is independent of the source term, and since we 
demand that the metric perturbation must vanish 
if the energy-momentum-stress tensor is zero (in 
which case, the spacetime must be a Minkowski space), 
then the complementary must vanish, which means 
that $\varepsilon_{\mu\nu} = 0$. 

As for the second part, it is the particular 
solution, which gives the source on the 
right-hand side of Eq. (\ref{2.19}) upon being 
acted on by $\Box$. Since the complementary 
solution is already zero, we simply write it 
as $\tilde{h}_{\mu\nu}$. This solution can be 
obtained by using the method of Green's function 
as follows. Write 
\begin{equation}
\tilde{h}_{\mu\nu}(x) = -\frac{16\pi G}{c^4}\int 
d^4y \, G(x-y)T_{\mu\nu}(y),               \label{2.21}
\end{equation}
where $G(x-y)$ is the Green's function satisfying 
\begin{equation}
\Box_x G(x-y) = \delta^4 (x-y).            \label{2.22}
\end{equation}
The solution (\ref{2.21}) clearly satisfies 
Eq. (\ref{2.19}), and so the problem of solving 
Eq. (\ref{2.19}) reduces to solving for the Green's 
function, which we do now. Roughly speaking, the 
Green's function in Eq. (\ref{2.22}) is the 
inverse of the linear differential operator 
$\Box$, satisfying certain boundary (or initial) 
conditions. Since Eq. (\ref{2.21}) says that 
the value of $\tilde{h}_{\mu\nu}$ at point 
$x^\mu$ depends on the value of $T_{\mu\nu}$ 
at point $y^\mu$ through $G(x-y)$, causality 
demands that $G(x-y) \ne 0$ only when $x^0 > y^0$, 
which comes from the requirement that the value 
of the metric perturbation at any time must 
depend on the source in its past (but not its 
future). Thus, a suitable condition that must 
be imposed on the Green's function is
\begin{equation}
G(x-y) = 0 \hspace{1cm} \mbox{when $x^0 < y^0$} 
                                          \label{2.23}
\end{equation}
and the resulting Green's function is called 
the retarded Green's function. 

Let us now find the Green's function satisfying 
the condition (\ref{2.23}). Motivated by 
\begin{equation}
\delta^4(x-y) = \frac{1}{(2\pi)^4}\int d^4k\,
e^{-ik \cdot (x-y)},                     \label{2.24}
\end{equation}
we write 
\begin{equation}
G(x-y) = \frac{1}{(2\pi)^4}\int d^4k\,
\tilde{G}(k)\, e^{-ik \cdot (x-y)}.      \label{2.25}
\end{equation}
Substituting (\ref{2.24}) and (\ref{2.25}) 
into Eq. (\ref{2.22}), we find 
\begin{equation}
\tilde{G}(k) = -\frac{1}{k^2},            \label{2.26}
\end{equation}
so that
\begin{equation}
G(x-y) = -\frac{1}{(2\pi)^4}\int d^4k\,
\frac{e^{-ik \cdot (x-y)}}{k^2}.        \label{2.27}
\end{equation}
To evaluate the above integral, we write $d^4k = 
dk^0d^3\vec{k}$ and $k^2 = (k^0)^2 - |\vec{k}|^2 
= (k^0-|\vec{k}|)(k^0+|\vec{k}|)$, so that
\begin{equation}
G(x-y) = -\frac{1}{(2\pi)^4} \int d^3\vec{k}\, 
e^{i\vec{k} \cdot (\vec{x}-\vec{y})} 
\int_{-\infty}^\infty dk^0\,\frac{e^{-ik^0(x^0-y^0)}}
{(k^0-|\vec{k}|)(k^0+|\vec{k}|)}.         \label{2.28}
\end{equation}
Let us now evaluate the integral over $k^0$. 
As the integrand has two simple poles at 
$k^0=\pm |\vec{k}|$, a direct integration on 
the real line of the complex $k^0$-plane is 
problematic. This problem, however, can be 
cured by deforming a contour a little bit in 
such a way that $k^0$ acquires a negligible 
imaginary part when passing the poles. Such 
a deformation clearly does not affect the 
fact that $G(x-y)$ in Eq. (\ref{2.27}) is a 
solution to Eq. (\ref{2.22}) as the operation 
of $\Box_x$ on $G(x-y)$ will remove these poles 
and will eventually give the delta function. 
There are many ways to deform the contour 
depending on the condition that we want to 
impose on $G(x-y)$. Due to the fact that the 
integrand in Eq. (\ref{2.28}) contains a 
factor $\exp(-ik^0(x^0-y^0))$, the requirement 
that $G(x-y)|_{x^0 < y^0} = 0$ forces the 
deformed contour to lie a little bit above the 
real line on the complex $k^0$-plane, as shown 
in Fig. 2.1. With such a contour, when 
$x^0 < y^0$ (or $(x^0 - y^0) < 0$), we have 
to close the contour on the upper-half plane 
in the evaluation of the integral (due to the 
fact that $\exp(-ik^0(x^0-y^0))$ is zero at 
infinity on the upper-half plane) and this 
gives zero as there is no pole inside the 
closed contour. Thus, the condition 
$G(x-y)|_{x^0 < y^0} = 0$ is satisfied. 


\begin{figure}[htbp]
\centerline{\includegraphics[scale=.5]{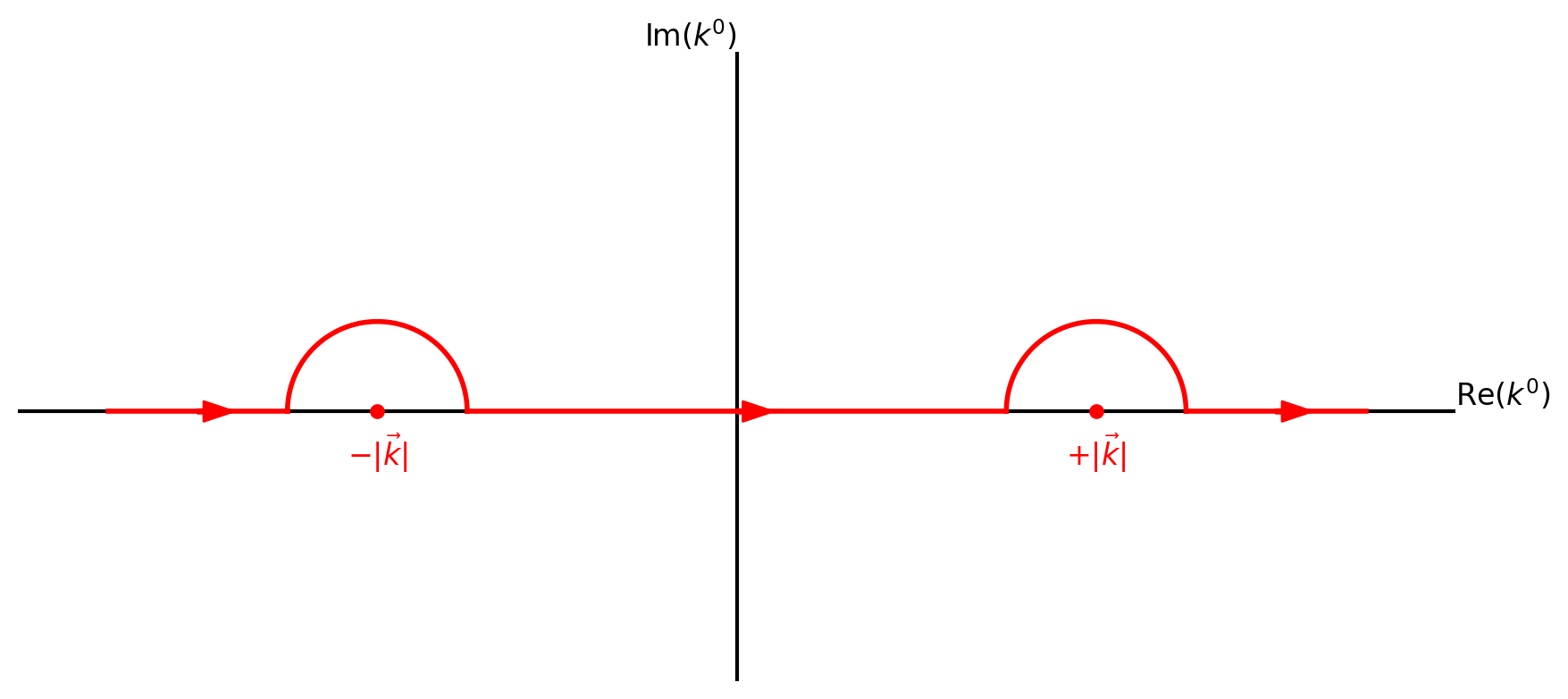}}
\caption{A deformed contour on 
the complex $k^0$-plane for the evaluation 
of the retarded Green's function. }
\label{fig}
\end{figure}


\vspace{1cm}

On the other hand, if $x^0 > y^0$ (or $(x^0 - 
y^0) > 0$), we must close the contour on the 
lower-half plane, which will enclose both poles. 
As this contour is clockwise, the result of 
this integration is equal to $(-2\pi i)$ times 
the sum over residues at both poles. Thus 
\begin{eqnarray}
G(x-y)|_{x^0 > y^0} &=& -\frac{1}{(2\pi)^4} 
\int d^3\vec{k}\, e^{i\vec{k} \cdot (\vec{x}-\vec{y})}\, 
(-2\pi i)\left( \frac{e^{-i|\vec{k}|(x^0-y^0)}}{2|\vec{k}|}
+ \frac{e^{i|\vec{k}|(x^0-y^0)}}{(-2|\vec{k}|)} \right) 
\nonumber \\
&=& \frac{1}{(2\pi)^3} \int d^3\vec{k}\, 
e^{i\vec{k} \cdot (\vec{x}-\vec{y})}\, 
\frac{\sin(|\vec{k}|(x^0-y^0))}{|\vec{k}|}.  \label{2.29}
\end{eqnarray}
We next evaluate the integral over $\vec{k}$. 
For convenience, we use a spherical coordinate 
system in which the $k^3$-axis is chosen to 
lie along the $\vec{r}\equiv (\vec{x}-\vec{y})$ 
direction. Write $\kappa\equiv |\vec{k}|$ and 
$\tau\equiv (x^0-y^0) > 0$, then $d^3\vec{k} = 
\kappa^2 d\kappa d(\cos\theta)d\phi$, 
$\vec{k}\!\cdot\!(\vec{x}-\vec{y}) = \kappa 
r\cos\theta$, $|\vec{k}|(x^0-y^0) = \kappa\tau$, 
and the integral (\ref{2.29}) can easily be 
integrated over $\cos\theta$ and $\phi$ to obtain 
\begin{eqnarray}
G(x-y)|_{x^0 > y^0} &=& \frac{1}{8\pi^2 r}
\int_0^\infty d\kappa \, \left[\left(
e^{i\kappa (r-\tau)}+e^{-i\kappa (r-\tau)}\right)
-\left( e^{i\kappa (r+\tau)}+e^{-i\kappa (r+\tau)} 
\right)\right]  \nonumber \\
&=& \frac{1}{8\pi^2 r}\int_{-\infty}^\infty d\kappa 
\,\left(e^{i\kappa (r-\tau)}-e^{i\kappa (r+\tau)} 
\right)  \nonumber \\
&=& \frac{1}{4\pi r}\left(\delta(r-\tau) - 
\delta(r+\tau)\right).                      \label{2.30}
\end{eqnarray}
Since $\tau\equiv (x^0-y^0) > 0$, then the 
second delta function in Eq. (\ref{2.30}) 
vanishes automatically. We thus obtain the 
retarded Green's function of the operator $\Box$,
\begin{equation}
G(x-y) = \frac{1}{4\pi |\vec{x}-\vec{y}|}
\delta(|\vec{x}-\vec{y}|-(x^0-y^0)),        \label{2.31}
\end{equation}
where we have dropped $|_{x^0 > y^0}$ since the 
delta function vanishes when $x^0 < y^0$ anyway. 
Substituting $G(x-y)$ in Eq. (\ref{2.31}) into 
Eq. (\ref{2.21}) and performing the integration 
over $y^0$, we finally obtain
\begin{equation}
\tilde{h}_{\mu\nu}(x^0,\vec{x}) = -\frac{4G}{c^4}
\int d^3y \, \frac{1}{|\vec{x}-\vec{y}|}\, 
T_{\mu\nu}(x^0-|\vec{x}-\vec{y}|,\vec{y}).  \label{2.32}
\end{equation}
Eq. (\ref{2.32}) describes $\tilde{h}_{\mu\nu}(x)$ 
as resulting from the source $T_{\mu\nu}(y)$ at 
all spacetime points $y$ on the past light cone 
of $x$ as shown in Fig. 2.2. In the next section, 
we will calculate $\tilde{h}_{\mu\nu}$ when the 
matter has no pressure and shear stress. 


\begin{figure}[htbp]
\centerline{\includegraphics[scale=.5]{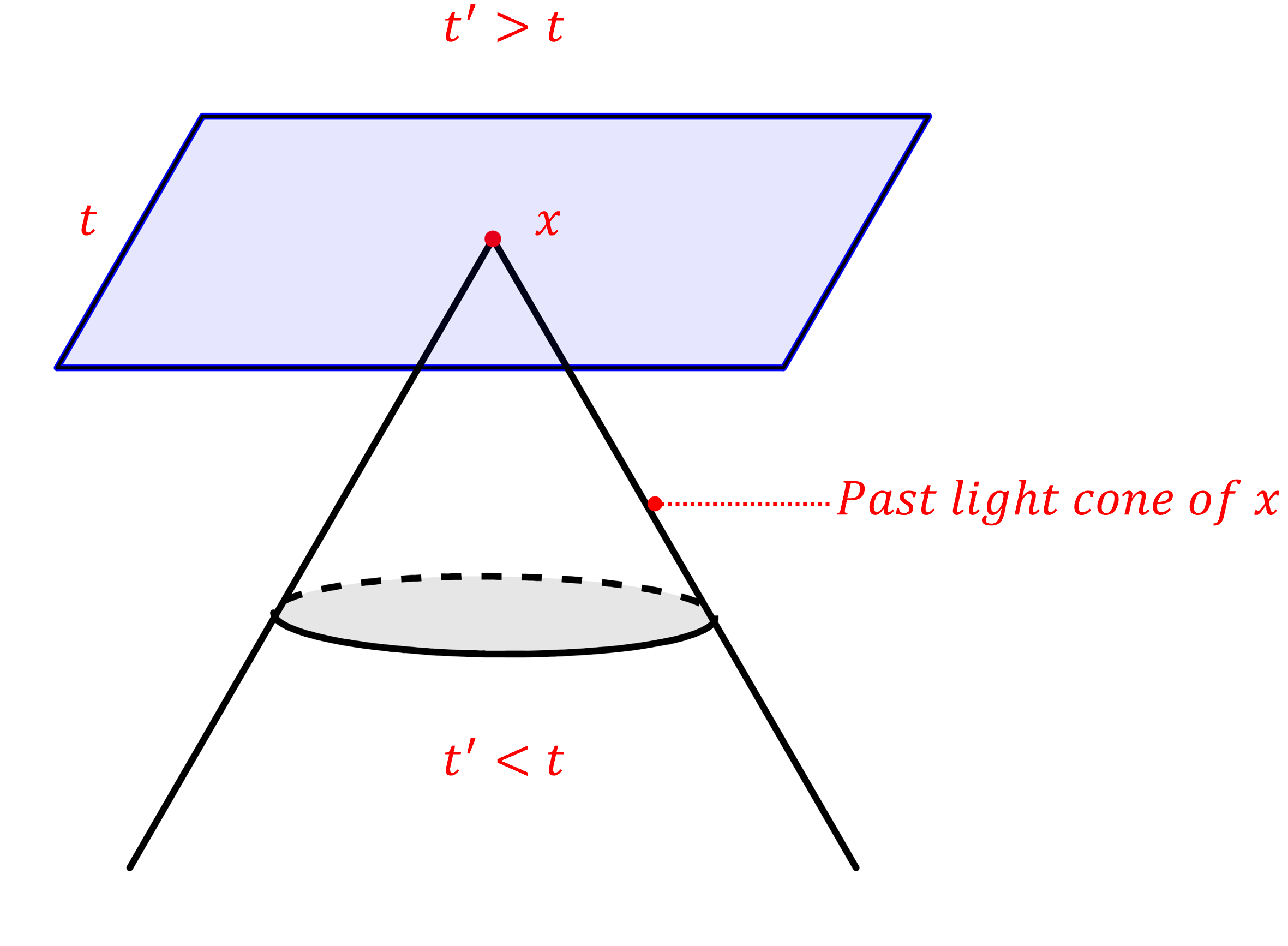}}
\caption{$\tilde{h}_{\mu\nu}(x)$ as 
due to the source on the past light cone of $x$. }
\label{fig}
\end{figure}




\section{Matter without Pressure and Shear Stress}

We now consider a specific example of matter with 
no pressure and shear stress. This means that the 
$T^{ij}$ components of the energy-momentum-stress 
tensor vanish, and so $T^{\mu\nu}$ takes the form
\begin{equation}
T^{\mu\nu} = \left( \begin{array}{cccc}
                    \rho c^2 & cj^1 & cj^2 & cj^3 \\
                    cj^1 & 0 & 0 & 0 \\
                    cj^2 & 0 & 0 & 0 \\
                    cj^3 & 0 & 0 & 0 
                    \end{array} \right)     \label{2.33}
\end{equation}
where $\rho$ and $j^i$ are, respectively, the mass 
density and the momentum density, and $T_{\mu\nu}$ 
is obtained from $T^{\mu\nu}$ by lowering the indices 
using $\eta_{\mu\nu}$. Substituting the above 
$T_{\mu\nu}$ into Eq. (\ref{2.32}), we obtain 
the non-zero components of $\tilde{h}_{\mu\nu}$, 
\begin{eqnarray}
\tilde{h}_{00}(ct,\vec{x}) &=& -\frac{4G}{c^2}
\int d^3y \, \frac{1}{|\vec{x}-\vec{y}|}\, 
\rho(ct-|\vec{x}-\vec{y}|,\vec{y})          \label{2.34} \\
\tilde{h}_{0i}(ct,\vec{x}) &=& -\frac{4G}{c^3}
\int d^3y \, \frac{1}{|\vec{x}-\vec{y}|}\, 
j_i(ct-|\vec{x}-\vec{y}|,\vec{y}),          \label{2.35} 
\end{eqnarray}
where $j_i = -j^i$, and we have written $x^0=ct$ 
explicitly. Let
\begin{eqnarray}
\Phi(ct,\vec{x}) &\equiv & \frac{G}{c^2}
\int d^3y \, \frac{1}{|\vec{x}-\vec{y}|}\, 
\rho(ct-|\vec{x}-\vec{y}|,\vec{y})          \label{2.36} \\
A_i(ct,\vec{x}) &\equiv & \frac{2G}{c^3}
\int d^3y \, \frac{1}{|\vec{x}-\vec{y}|}\, 
j_i(ct-|\vec{x}-\vec{y}|,\vec{y}),          \label{2.37}
\end{eqnarray}
then $\tilde{h}_{00} = -4\Phi$ and $\tilde{h}_{0i} 
= -2A_i$. From the definition of $\tilde{h}_{\mu\nu}$ 
in Eq. (\ref{2.18}), 
\begin{equation}
\tilde{h}_{\mu\nu} \equiv h_{\mu\nu} - 
\frac{\tilde{\lambda}}{2}\eta_{\mu\nu}h,    \label{2.38}
\end{equation}
and the above form of $\tilde{h}_{\mu\nu}$, we find 
$\tilde{h} \equiv \eta^{\mu\nu}\tilde{h}_{\mu\nu} = 
(1-2\tilde{\lambda})h = -4\Phi$. Thus 
\begin{equation}
h = -\frac{4\Phi}{(1-2\tilde{\lambda})},    \label{2.39}
\end{equation}
so that
\begin{eqnarray}
h_{00} &=& \tilde{h}_{00} + \frac{\tilde{\lambda}}{2}
\eta_{00}h  \nonumber \\
&=& -2\left(\frac{2-3\tilde{\lambda}}{1-2\tilde{\lambda}}
\right)\Phi                                 \label{2.40} \\
h_{0i} &=& \tilde{h}_{0i} + \frac{\tilde{\lambda}}{2}
\eta_{0i}h  \nonumber \\
&=& -2A_i                                   \label{2.41} \\
h_{ij} &=& \tilde{h}_{ij} + \frac{\tilde{\lambda}}{2}
\eta_{ij}h  \nonumber \\
&=& \left(\frac{2\tilde{\lambda}\Phi}{1-2\tilde{\lambda}}
\right)\delta_{ij}.                         \label{2.42}
\end{eqnarray}
To express the above result in terms of the original 
parameter $\lambda$, we recall that $\tilde{\lambda} 
= 1-2\lambda$, so that Eqs. (\ref{2.40})--(\ref{2.42}) 
become 
\begin{eqnarray}
h_{00} &=& -2\left(\frac{1-6\lambda}{1-4\lambda}\right)
\Phi                                        \label{2.43} \\
h_{0i} &=& -2A_i                            \label{2.44} \\
h_{ij} &=& -2\left(\frac{1-2\lambda}{1-4\lambda}\right)
\Phi\delta_{ij}.                            \label{2.45}
\end{eqnarray}
We thus obtain the linearized metric corresponding 
to the energy-momentum-stress tensor (\ref{2.33}),
\begin{eqnarray}
g_{00} &=& 1-\alpha\Phi                     \label{2.46} \\
g_{0i} &=& -2A_i                            \label{2.47} \\
g_{ij} &=& -(1+\beta\Phi)\delta_{ij}        \label{2.48}
\end{eqnarray}
where we have defined 
\begin{eqnarray}
\alpha &\equiv & 
2\left(\frac{1-6\lambda}{1-4\lambda}\right)  \label{2.49} \\
\beta &\equiv & 
2\left(\frac{1-2\lambda}{1-4\lambda}\right). \label{2.50}
\end{eqnarray}
Thus, we finally obtain the linearized metric
\begin{equation}
ds^2 = (1-\alpha\Phi)c^2dt^2 - 4A_i\,cdtdx^i - 
(1+\beta\Phi)\delta_{ij}dx^idx^j.            \label{2.51}
\end{equation}

Despite the fact that $\Phi$ and $A_i$ in Eqs. 
(\ref{2.36}) and (\ref{2.37}) look complicated, 
their forms can be simplified enormously if we 
consider a special case in which the matter are 
localized in space and are slowly varying in time 
\cite{Jackson}. By slowly varying in time, we 
mean $\rho$ and $j_i$ can be approximately 
treated as time-independent, which leads us 
to write 
\begin{eqnarray}
\Phi(\vec{x}) &\approx & \frac{G}{c^2}
\int d^3y \, \frac{1}{|\vec{x}-\vec{y}|}\, 
\rho(\vec{y})                                \label{2.52} \\
A_i(\vec{x}) &\approx & \frac{2G}{c^3}
\int d^3y \, \frac{1}{|\vec{x}-\vec{y}|}\, 
j_i(\vec{y}).                                \label{2.53}
\end{eqnarray}
To satisfy the assumption of localization in space, 
we demand that $\rho(\vec{y})$ and $j_i(\vec{y})$ 
must tend to zero fast enough as $|\vec{y}|$ becomes 
large, which means that the matter is concentrated 
in a finite volume in space. This enables us to expand 
\begin{eqnarray}
\frac{1}{|\vec{x}-\vec{y}|} &=& \frac{1}
{\sqrt{(\vec{x}-\vec{y})\!\cdot\! (\vec{x}-\vec{y})}}  
\nonumber \\
&=& \frac{1}{\sqrt{|\vec{x}|^2 - 2\vec{x}\!\cdot\!\vec{y} 
+ |\vec{y}|^2}}  \nonumber \\ 
&=& \frac{1}{|\vec{x}|\sqrt{1 - 
(2\vec{x}\!\cdot\!\vec{y}/|\vec{x}|^2) + 
(|\vec{y}|^2/|\vec{x}|^2)}}  \nonumber \\ 
&=& \frac{1}{|\vec{x}|}\left(1 + 
\frac{\vec{x}\!\cdot\!\vec{y}}{|\vec{x}|^2} + 
\ldots \,\right)  \nonumber \\ 
&=& \frac{1}{|\vec{x}|} + 
\frac{\vec{x}\!\cdot\!\vec{y}}{|\vec{x}|^3} + \ldots ,
                                             \label{2.54}
\end{eqnarray}
in the above integrals if $|\vec{x}|$ is much 
larger than the localization length scale. Thus 
\begin{equation}
\Phi(\vec{x}) \approx \frac{G}{c^2}
\int d^3y \, \frac{1}{|\vec{x}|}\,\rho(\vec{y}) + 
\frac{G}{c^2}\int d^3y \, 
\frac{\vec{x}\!\cdot\!\vec{y}}{|\vec{x}|^3}\,
\rho(\vec{y}).                               \label{2.55}
\end{equation}
Note that the termination of the infinite series 
(\ref{2.54}) at the second term in Eq. (\ref{2.55}) 
perfectly makes sense, since the linearized metric 
is valid only in a weak gravitational field, i.e., 
when $|\vec{x}|$ is much larger than the length 
scale of the mass that mediates gravity. The 
second term in Eq. (\ref{2.55}) is zero if we 
choose the origin of the coordinate system to 
be at the center of mass. Thus Eq. (\ref{2.55}) 
simplifies to
\begin{equation}
\Phi(\vec{x}) \approx \frac{GM}{c^2|\vec{x}|}, 
                                             \label{2.56}
\end{equation}
where $M\equiv\int d^3y\,\rho(\vec{y})$ is the 
total mass. As for $A_i$, we do the same thing 
to obtain
\begin{equation}
A_i(\vec{x}) \approx \frac{2G}{c^3}
\int d^3y \, \frac{1}{|\vec{x}|}\,j_i(\vec{y}) + 
\frac{2G}{c^3}\int d^3y \, 
\frac{\vec{x}\!\cdot\!\vec{y}}{|\vec{x}|^3}\,
j_i(\vec{y}).                               \label{2.57}
\end{equation}
To evaluate the above integrals, we add one more 
assumption that the current $j_i$ forms a closed 
loop, which implies $\vec{\nabla}\!\cdot\!\vec{j} 
= \partial_i j^i = 0$. Now, consider the identity
\begin{eqnarray}
0 &=& \int d^3y\,\vec{\nabla}\!\cdot\!(fg\vec{j}) 
\nonumber \\
&=& \int d^3y\,\left(f(\vec{j}\!\cdot\!\vec{\nabla})g 
+ g(\vec{j}\!\cdot\!\vec{\nabla})f\right),  \label{2.58}
\end{eqnarray}
for some functions $f(\vec{y})$ and $g(\vec{y})$ 
such that the product $fg\vec{j}$ goes to zero 
faster than $1/|\vec{y}|^2$ as $|\vec{y}|\to\infty$ 
so as to make the surface term on the first line 
disappear. Above, we have used the property 
$\vec{\nabla}\!\cdot\!\vec{j} = 0$ on the second line. 
Choosing $f = 1$ and $g = y^i$, Eq. (\ref{2.58}) 
leads to 
\begin{equation}
\int d^3y \, j^i(\vec{y}) = 0,              \label{2.59}
\end{equation}
so that the first term in Eq. (\ref{2.57}) is zero. 
On the other hand, if we choose $f = y^i$ and $g = 
y^j$, then Eq. (\ref{2.58}) implies
\begin{equation}
\int d^3y \, \left(y^i j^j + y^j j^i\right) = 0. 
                                            \label{2.60}
\end{equation}
We can use this result to express the second 
integral in Eq. (\ref{2.57}) as 
\begin{eqnarray}
2\int d^3y \, 
(\vec{x}\!\cdot\!\vec{y})\,
j_i(\vec{y}) &=& 2\sum_{j=1}^3 x^j\int d^3y \, 
y^j j_i(\vec{y})  \nonumber \\
&=& -2\sum_{j=1}^3 x^j\int d^3y \, y^j j^i(\vec{y})
\nonumber \\
&=& -\sum_{j=1}^3 x^j\int d^3y \, \left(y^j j^i(\vec{y})
- y^i j^j(\vec{y})\right)  \nonumber \\
&=& \sum_{j,k=1}^3 \epsilon_{ijk}x^j\int d^3y \, 
(\vec{y}\times\vec{j})^k  \nonumber \\
&=& \left(\vec{x}\times\int d^3y \, 
(\vec{y}\times\vec{j})\right)^i  \nonumber \\
&=& \left(\vec{x}\times\vec{S}\right)^i,   \label{2.61}
\end{eqnarray}
where we have used the fact that the momentum 
density is $j^i = -j_i$, and we have defined 
$\vec{S}$ as the total angular momentum about 
the center of mass, 
\begin{equation}
\vec{S} \equiv \int d^3y \, (\vec{y}\times\vec{j}). 
                                            \label{2.62}
\end{equation}
Thus
\begin{equation}
A_i(\vec{x}) = \frac{G\left(\vec{x}\times\vec{S}\right)^i}
{c^3|\vec{x}|^3},                           \label{2.63}
\end{equation}
where the upper index $i$ on the right-hand 
side indicates that this quantity is a 3-dimensional 
vector in the usual sense. If $\vec{S}$ points 
along the $z$-direction in the Cartesian coordinate 
system, $\vec{S} = S\hat{z}$, then
\begin{eqnarray}
A_1(\vec{x}) &=& \frac{GyS}{c^3|\vec{x}|^3}   \label{2.64} \\
A_2(\vec{x}) &=& -\frac{GxS}{c^3|\vec{x}|^3}  \label{2.65} \\
A_3(\vec{x}) &=& 0.                           \label{2.66}
\end{eqnarray}
Substitute the results in Eq. (\ref{2.56}) and Eqs. 
(\ref{2.64})--(\ref{2.66}) into the metric 
(\ref{2.51}), we obtain
\begin{equation}
ds^2 = \left(1-\frac{\alpha GM}{c^2|\vec{x}|}
\right)c^2dt^2 - \left(1+\frac{\beta GM}{c^2|\vec{x}|}
\right)(dx^2+dy^2+dz^2) - \frac{4GS}{c^2|\vec{x}|^3}
(ydx-xdy)dt                                   \label{2.67}
\end{equation}
written explicitly in terms of the Cartesian 
coordinates. 

Consider a special case in which the mass $M$ 
is non-rotating and spherically symmetric, 
which means that the resulting gravitational 
field is spherically symmetric. In this case, 
$S=0$ and there must exist a coordinate system 
in which the spherical symmetry is manifest in 
the metric. To find such a coordinate system, 
we first rewrite the metric (\ref{2.67}) with 
$S=0$ in a spherical coordinate system,
\begin{equation}
ds^2 = \left(1-\frac{\alpha GM}{c^2r}\right)c^2dt^2 
- \left(1+\frac{\beta GM}{c^2r}\right)(dr^2 + 
r^2d\Omega^2),                               \label{2.68}
\end{equation}
where $r=|\vec{x}|$ and $d\Omega^2 = d\theta^2 + 
\sin^2\theta d\phi^2$. Our purpose is to write 
the metric (\ref{2.68}) in the form
\begin{equation}
ds^2 = \left(1-\frac{\alpha GM}{c^2r}\right)c^2dt^2 
- f(\bar{r})d\bar{r}^2 - \bar{r}^2d\Omega^2, \label{2.69}
\end{equation}
for some coordinate $\bar{r}$ and function 
$f(\bar{r})$. Compare Eq. (\ref{2.69}) with 
Eq. (\ref{2.68}), we find 
\begin{equation}
\bar{r} = \sqrt{1+\frac{\beta GM}{c^2r}}\,r. \label{2.70}
\end{equation}
Differentiating Eq. (\ref{2.70}), we find 
\begin{eqnarray}
d\bar{r} &=& \sqrt{1+\frac{\beta GM}{c^2r}}
\left(\frac{r+(\beta GM/2c^2)}{r+(\beta GM/c^2)}
\right)dr  \nonumber \\
&\approx & \sqrt{1+\frac{\beta GM}{c^2r}}
\left(1-\frac{1}{2}\frac{\beta GM}{c^2r}\right)dr, 
                                            \label{2.71}
\end{eqnarray}
valid up to the first order in $\Phi=GM/c^2r$. 
The first line of Eq. (\ref{2.71}) tells us 
that $d\bar{r}/dr > 0$, which means that 
$\bar{r}$ is an increasing function of $r$. 
Thus Eq. (\ref{2.70}) represents a good 
transformation from $r\in [0,\infty)$ to 
$\bar{r}\in [0,\infty)$. Also,
\begin{eqnarray}
\frac{GM}{c^2r} &=& \frac{GM}{c^2\bar{r}}
\sqrt{1+\frac{\beta GM}{c^2r}}  \nonumber \\
&\approx & \frac{GM}{c^2\bar{r}},           \label{2.72}
\end{eqnarray}
to the first order in $\Phi=GM/c^2r$. 
Combine Eqs. (\ref{2.71}) and (\ref{2.72}) 
together, we get
\begin{eqnarray}
\left(1+\frac{\beta GM}{c^2r}\right)dr^2 &=& 
\frac{d\bar{r}^2}{\left(1-\frac{1}{2}
\frac{\beta GM}{c^2\bar{r}}\right)^2}  \nonumber \\
&\approx & \left(1+\frac{\beta GM}{c^2\bar{r}}
\right)d\bar{r}^2,                         \label{2.73}
\end{eqnarray}
where we have discarded terms of the order 
$\Phi^2$ in all steps of the calculation. 
Using this result in Eq. (\ref{2.68}), we 
finally obtain 
\begin{equation}
ds^2 = \left(1-\frac{\alpha GM}{c^2\bar{r}}\right)c^2dt^2 
- \left(1+\frac{\beta GM}{c^2\bar{r}}\right)
d\bar{r}^2 - \bar{r}^2d\Omega^2,          \label{2.74}
\end{equation}
where we have used Eq. (\ref{2.72}) to replace 
$r$ by $\bar{r}$ in the first term. By treating 
$\bar{r}$ as the physical radial coordinate, 
the metric (\ref{2.74}) is a linearized 
spherically symmetric metric of Rastall gravity, 
which reduces to the linearized Schwarzschild metric in 
General Relativity in the limit $\lambda = 0$ 
(or, equivalently, $\alpha = \beta = 2$) as 
expected. We will use this metric in the next 
chapter when we discuss a phenomenological 
consequence of the linearized Rastall gravity.




\chapter{Phenomenological Consequence: Gravitoelectromagnetism}

Having obtained the linearized solution to Rastall gravity, we are 
now ready to discuss one of its phenomenological consequences, known 
as the gravitoelectromagnetism. The main idea is as follows. Consider 
a free-falling observer moving along a timelike geodesic in spacetime. 
Suppose this observer performs an experiment which involves a 
freely-moving object (such as a spinning top) in his co-moving 
reference frame, it is interesting to investigate the law of 
motion satisfied by the object as seen by the free-falling 
observer. As will be seen in this chapter, it turns out that 
to the lowest-order approximation in the object's speed relative 
to the observer divided by the speed of light, this observer 
will see that the object nearby him experiences two kinds of 
forces. The first one is independent of the object's speed, 
while the second one is linearly proportional to the object's 
speed and mimics the form of the Lorentz force in electromagnetism. 
These two forces are, respectively, known as the gravitoelectric 
and gravitomagnetic forces \cite{Mashhoon1,Mashhoon2}. 

The organization of the chapter is as follows. In the first section, 
we will present the mathematical formulation of the Fermi normal 
coordinate system used by the free-falling observer in detail. We 
then go on to obtain the law of motion in the Fermi normal coordinate 
system in the second section, and see how the gravitoelectric and 
gravitomagnetic forces arise. In the last section, we will compute 
the gravitoelectric and gravitomagnetic fields as seen by an observer 
moving along a circular orbit in a weak gravitational field described 
by the linearized metric obtained in Chapter 2. As will be seen, 
the resulting gravitomagnetic field is perpendicular to the moving 
direction of the observer, which is quite an interesting result. 


\section{Fermi Normal Coordinate System}

In this section, we will define the Fermi normal coordinates in 
the region around a timelike geodesic (treated as the worldline 
of a free-falling observer), and then go on to obtain the 
corresponding metric tensor. We will follow closely the 
presentation in \cite{Poisson}, and use the unit $c=1$ for 
convenience. 


\subsection{Definition of the Fermi Normal Coordinates}

Consider a timelike geodesic $\gamma$ with the affine parameter 
$\tau$ and tangent vector $u = \partial/\partial\tau$ as shown 
in Fig. 3.1. (Recall that the affine parameter is the curve 
parameter such that the length squared of the tangent vector 
is constant along the curve.) 


\begin{figure}[htbp]
\centerline{\includegraphics[scale=.5]{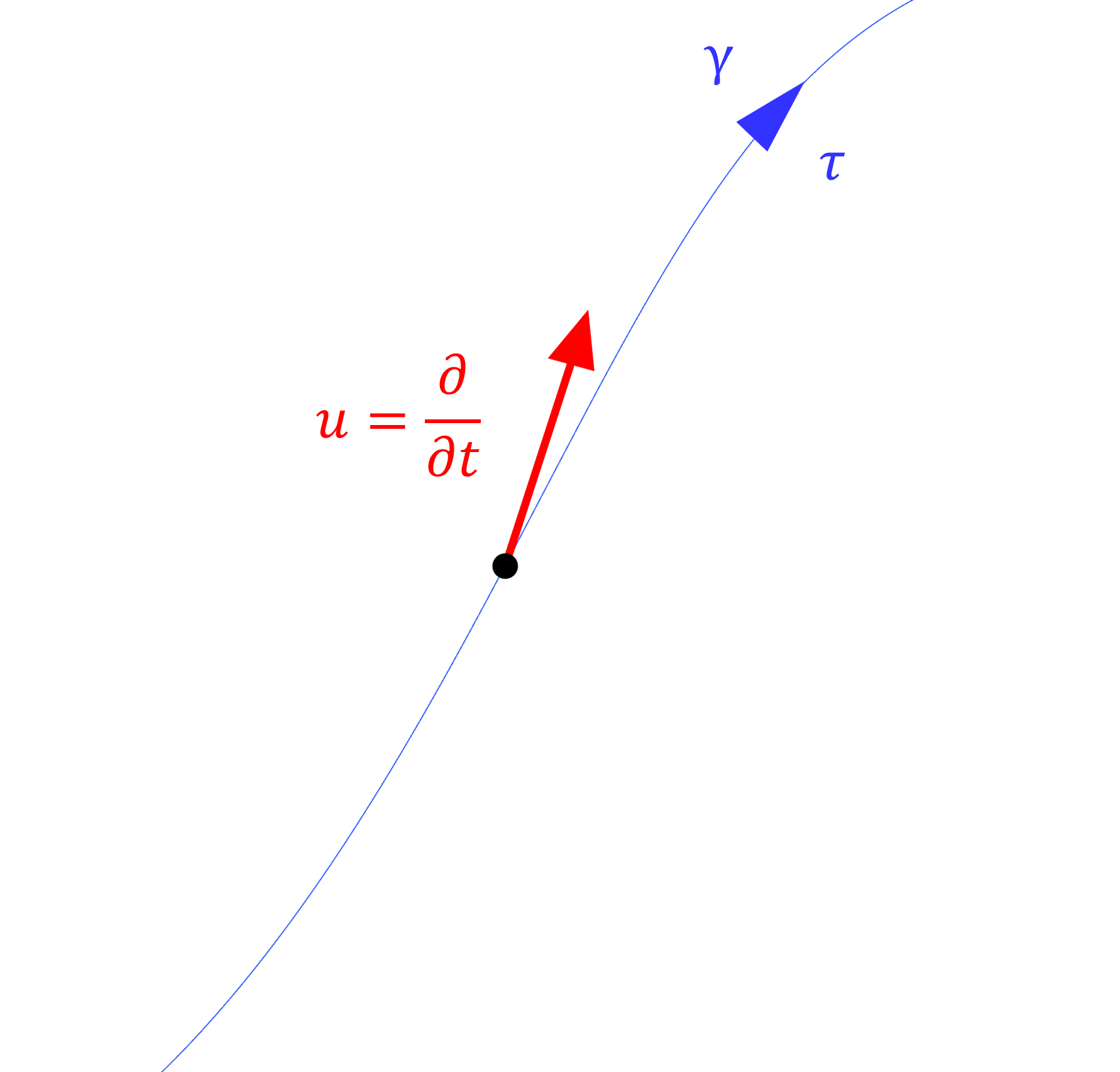}}
\caption{A timelike geodesic $\gamma$ and the 
corresponding tangent vector, $u=\partial/\partial\tau$.}
\label{fig}
\end{figure}


\vspace{1cm}

\noindent We can construct a set of vierbein $e_a$ $(a = 0,1,2,3)$ 
satisfying $e_a\!\cdot\! e_b \equiv g_{\mu\nu}e_a^\mu e_b^\nu = 
\eta_{ab}$ (with $\eta_{ab}$ being the Minkowski metric) at every 
point on $\gamma$ as follows. Let $O$ be a point on $\gamma$, which 
may be chosen to be the point $\tau =0$. We define the vierbein at $O$ 
by setting $e_0|_O \equiv u|_{\tau =0}$ and choosing $e_1|_O$, $e_2|_O$ 
and $e_3|_O$ to be a set of mutually orthogonal spacelike vectors of 
unit length perpendicular to $e_0|_O$ as shown in Fig. 3.2 (a). 
Clearly, $e_a|_O\!\cdot\! e_b|_O = \eta_{ab}$ by construction. The 
set of vierbein at any other point on $\gamma$ is subsequently 
constructed by parallel transporting $e_a|_O$ along $\gamma$ to the 
point of interest as in Fig. 3.2 (b). As the geodesic is the curve 
on which its tangent vector is parallel transported along itself, 
it is clear that $e_0 = u$ at all points on $\gamma$. As for 
$e_i$ $(i=1,2,3)$, they can in principle be found by solving 
the equation
\begin{equation}
u^\mu\nabla_\mu e_i^\nu = 0 
\end{equation}
describing the parallel transportation of $e_i$ along $\gamma$, 
with the initial condition $e_i|_{\tau =0} = e_i|_O$. Thus, the 
vierbein on $\gamma$ satisfies
\begin{equation}
u^\mu\nabla_\mu e_a^\nu = 0  \hspace{1cm} \mbox{$a=0,\ldots ,3$.}
                                                  \label{3.0.0}
\end{equation}
Since the parallel transportation preserves the inner product of 
vectors, the vierbein field constructed in this way automatically 
satisfies $e_a\!\cdot\! e_b = \eta_{ab}$ at all points on $\gamma$. 


\begin{figure}[htbp]
\centerline{\includegraphics[scale=.5]{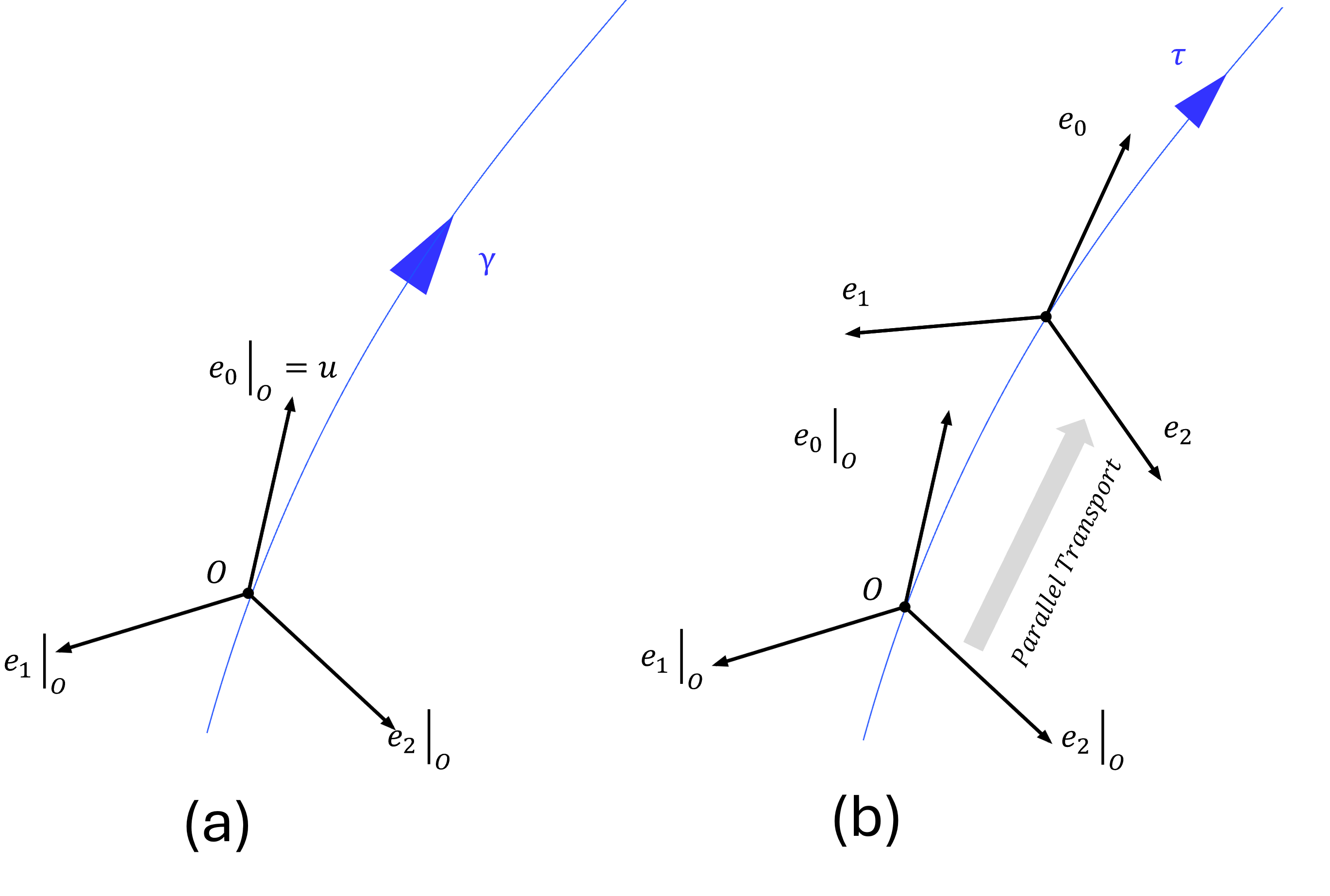}}
\caption{(a) The vierbein at the point $\tau =0$. 
(b) The construction of the vierbein on the timelike geodesic by 
parallel transportation.}
\label{fig}
\end{figure}


\newpage

Let us now consider a spacelike unit vector $v$ perpendicular to 
the timelike geodesic $\gamma$ at an arbitrary point $P$ as shown 
in Fig. 3.3 (a). Since the set $\{e_i|_P|i=1,2,3\}$ serves as an 
orthonormal basis of the 3-dimensional space perpendicular to 
$\gamma$ at point $P$, we can write
\begin{equation}
v^\mu = \Omega^i e_i^\mu                           \label{3.0}
\end{equation}
with the parameters $\Omega^i$ $(i=1,2,3)$ satisfying $\eta_{ij}
\Omega^i\Omega^j = -1$ so as to make $v^2 = -1$. Here, we have 
omitted the notation $|_P$ for simplicity. It is easy to see 
that the set of parameters $\Omega^i$ determines the direction of 
$v$, which implies the one-to-one correspondence between ``the 
set of $\Omega^i$" and ``the direction of $v$." By treating $v$ 
as the tangent vector at the initial point of a spacelike geodesic 
emanating from $P$ as shown in Fig. 3.3 (b), we can in principle 
solve the geodesic equation 
\begin{equation}
\frac{d^2x^\mu}{ds^2} + \Gamma^\mu_{\nu\rho}\frac{dx^\nu}{ds}
\frac{dx^\rho}{ds} = 0                             \label{3.0.1}
\end{equation}
with $s$ being the affine parameter, subject to the initial condition 
\begin{equation}
\left.\frac{dx^\mu}{ds}\right|_{s=0} = v^\mu ,     \label{3.0.2}
\end{equation}
to obtain such a spacelike geodesic. 


\begin{figure}[htbp]
\centerline{\includegraphics[scale=.5]{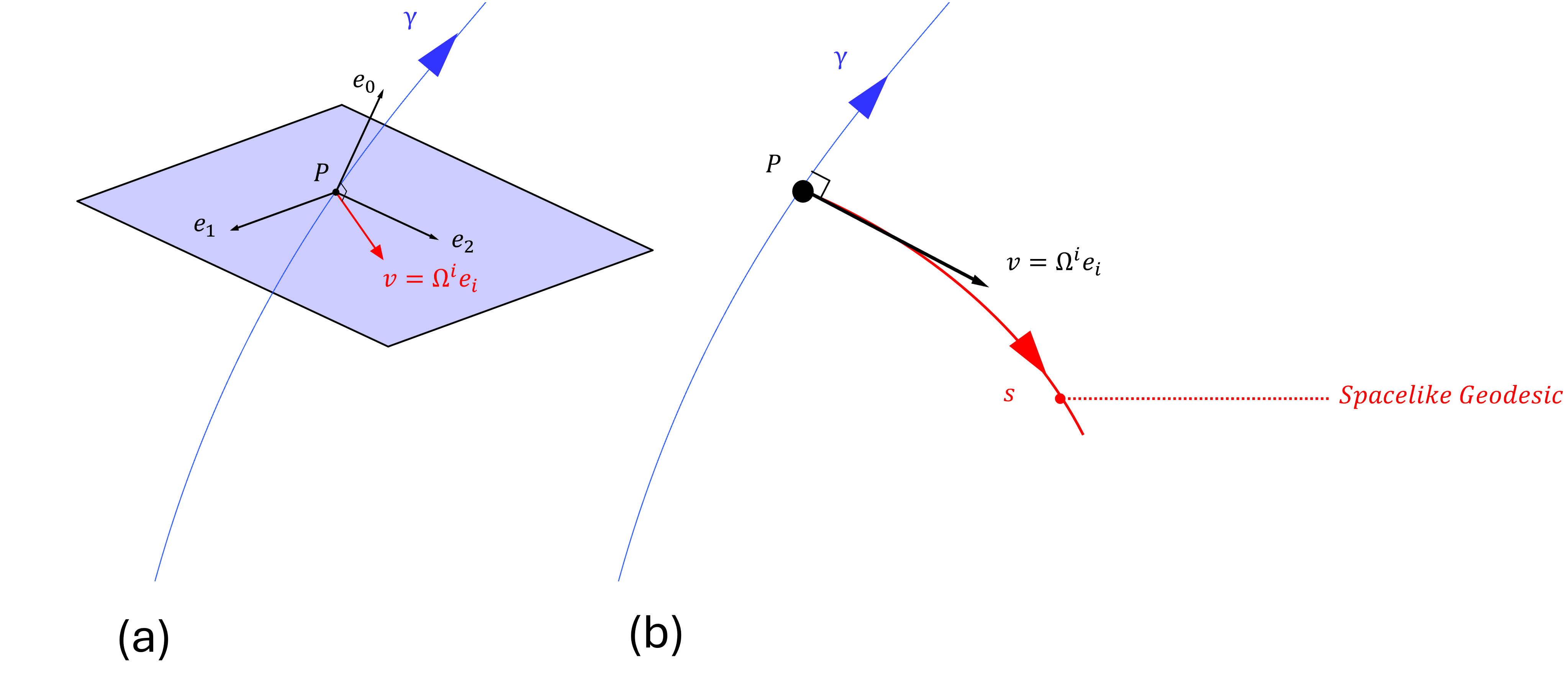}}
\caption{(a) A spacelike unit vector $v^\mu$ at point 
$P$. (b) A spacelike geodesic emanating from $P$ in the direction 
of $v^\mu$. }
\label{fig}
\end{figure}


\newpage

Let $Q$ be a point on this spacelike geodesic, and let $s_Q$ be the 
proper distance between $P$ and $Q$ along this geodesic. We define 
the ``Fermi normal coordinates" at point $Q$ by
\begin{equation}
\left. X^0 \right|_Q \equiv \tau_P \hspace{0.5cm} \mbox{and} 
\hspace{0.5cm} \left. X^i \right|_Q \equiv \Omega^i_Q s_Q,    \label{3.1}
\end{equation}
where $\tau_P$ is the affine parameter $\tau$ at point $P$, and we 
have added a subscript $Q$ to $\Omega^i$ to indicate that the vector 
$v^\mu_Q \equiv \Omega^i_Q e_i^\mu$ is the tangent vector at the 
initial point of the spacelike geodesic passing through $Q$. 


\begin{figure}[htbp]
\centerline{\includegraphics[scale=.5]{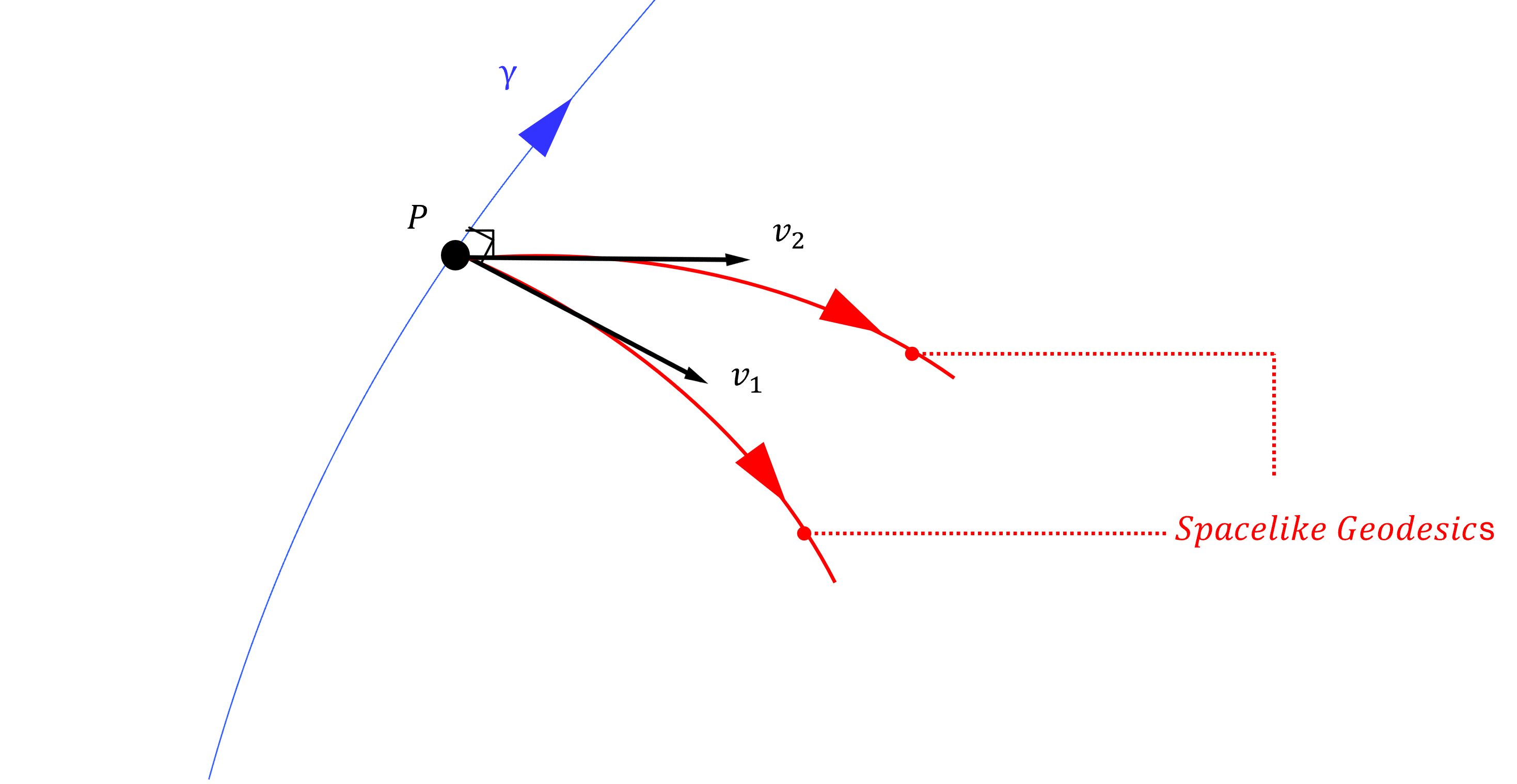}}
\caption{Two spacelike geodesics emanating from the 
same point on the timelike geodesic in two different directions. }
\label{fig}
\end{figure}


\newpage

At this point, an important caution should be made. Suppose $v_1$ 
and $v_2$ are different spacelike unit vectors at point $P$ as 
shown in Fig. 3.4. Since $v_1$ and $v_2$ are of different directions, 
each of them serves as the initial tangent vector to a different 
spacelike geodesic. In general, there is nothing to guarantee that 
these two spacelike geodesics will not cross each other; this might 
happen if the spacetime curvature is large enough in the vicinity 
of $\gamma$. If these geodesics cross at some point, say $\tilde{Q}$, 
then Eq. (\ref{3.1}) tells us that the definition of the Fermi 
normal coordinates at point $\tilde{Q}$ is not well-defined. But 
if the spacetime curvature is not too large in a sufficiently-small 
vicinity of the timelike geodesic $\gamma$, it is reasonable to 
claim that any two spacelike geodesics emanating from $\gamma$ 
will not cross each other. This ensures us that the Fermi normal 
coordinates in a small vicinity of $\gamma$ are well defined, 
especially in the weak-gravity region in which the linearized 
treatment holds. This is why the formalism of Fermi normal 
coordinates is suitable for the analysis of linearized gravity. 

The importance of the Fermi normal coordinates is as follows. 
Consider an astronaut sitting inside a spaceship orbiting around 
the Earth in a free-falling manner. The worldline of this astronaut 
is clearly the timelike geodesic $\gamma$. Suppose the spaceship 
is sufficiently small that any spacelike geodesic emanating from 
$\gamma$ is approximately a straight line inside the spaceship. 
Then at any instant, the astronaut can use the vierbein $e_a$ 
as the basis vectors in his co-moving reference frame, and that 
his reference frame is locally a Minkowski space. Since $\Omega^i$ 
in Eq. (\ref{3.0}) satisfies $\eta_{ij}\Omega^i\Omega^j = 
-\delta_{ij}\Omega^i\Omega^j = -1$, we can think of $\vec{\Omega} 
\equiv (\Omega^1,\Omega^2,\Omega^3)$ as a unit vector in a 
3-dimensional Euclidean space spanned by $(e_1,e_2,e_3)$. This 
means that the Fermi normal coordinates $\left.X^i \right|_Q = 
\Omega^i_Q s_Q$ at point $Q$ inside the spaceship are just the 
Euclidean components of a position vector $\vec{X} \equiv 
(X^1,X^2,X^3)$ pointing from the astronaut to point $Q$ nearby 
him. This interpretation allows us to express the location of a 
thing inside the spaceship in terms of the Fermi normal coordinates. 
In other words, in a small region nearby the timelike geodesic 
$\gamma$, the Fermi normal coordinate system is just a local 
Cartesian coordinate system that the astronaut uses to pinpoint 
the location of any object nearby him.


\subsection{Link Between the Spacetime Coordinates and the Fermi 
Normal Coordinates}

From the above interpretation of the Fermi normal coordinates and 
the fact that we normally calculate things in terms of the spacetime 
coordinates, it is important to find the link between the spacetime 
coordinates $x^\mu$ and the Fermi normal coordinates $X^a$ as this 
link will enable us to convert any tensorial quantity calculated in 
the coordinate basis to the corresponding quantity expressed in the 
Fermi normal coordinate system, which is precisely the quantity 
that the astronaut measures inside his spaceship. To obtain such 
a link, we first note that the vector $v^\mu = \Omega^i e_i^\mu$ in 
Eq. (\ref{3.0}), being treated as the tangent vector of a spacelike 
geodesic at point $P$, can be written as (see Eq. (\ref{3.0.2}))
\begin{equation}
v^\mu = \left.\frac{dx^\mu}{ds}\right|_{\tau_P,\Omega^i}     \label{3.2}
\end{equation}
where $s$ is the affine parameter of the spacelike geodesic, $\tau_P$ 
is the affine parameter $\tau$ at point $P$ of the timelike geodesic, 
and $\Omega^i$ indicates the direction of the spacelike geodesic as 
it emanates from point $P$. As mentioned before, we can obtain the 
spacelike geodesic by solving the geodesic equation
\begin{equation}
\frac{d^2x^\mu}{ds^2} + \Gamma_{\nu\rho}^\mu\frac{dx^\nu}{ds}
\frac{dx^\rho}{ds} = 0                                   \label{3.3}
\end{equation}
subject to the initial condition
\begin{equation}
\left.\frac{dx^\mu}{ds}\right|_{s=0} = v^\mu .           \label{3.4}
\end{equation}
By performing a rescaling of the affine parameter, $s\rightarrow s/c$ 
(with $c$ being a positive constant), the geodesic equation (\ref{3.3}) 
still takes the same form, but the initial condition (\ref{3.4}) 
changes to
\begin{equation}
\left.\frac{dx^\mu}{d(s/c)}\right|_{s=0} = cv^\mu ,      \label{3.5}
\end{equation}
which amounts to changing $\Omega^i \rightarrow c\Omega^i$. Since the 
location of any point on the spacelike geodesic depends on (i) 
the affine parameter $\tau$ of the timelike geodesic $\gamma$ at its 
initial point, (ii) the direction $\Omega^i$ at its starting point, 
and (iii) the value of the affine parameter $s$ on the spacelike 
geodesic, we may write the coordinate $x^\mu$ on the spacelike 
geodesic as
\begin{equation}
x^\mu = x^\mu(\tau,\Omega^i,s)                              \label{3.6}
\end{equation}
as shown in Fig. 3.5.


\begin{figure}[htbp]
\centerline{\includegraphics[scale=.5]{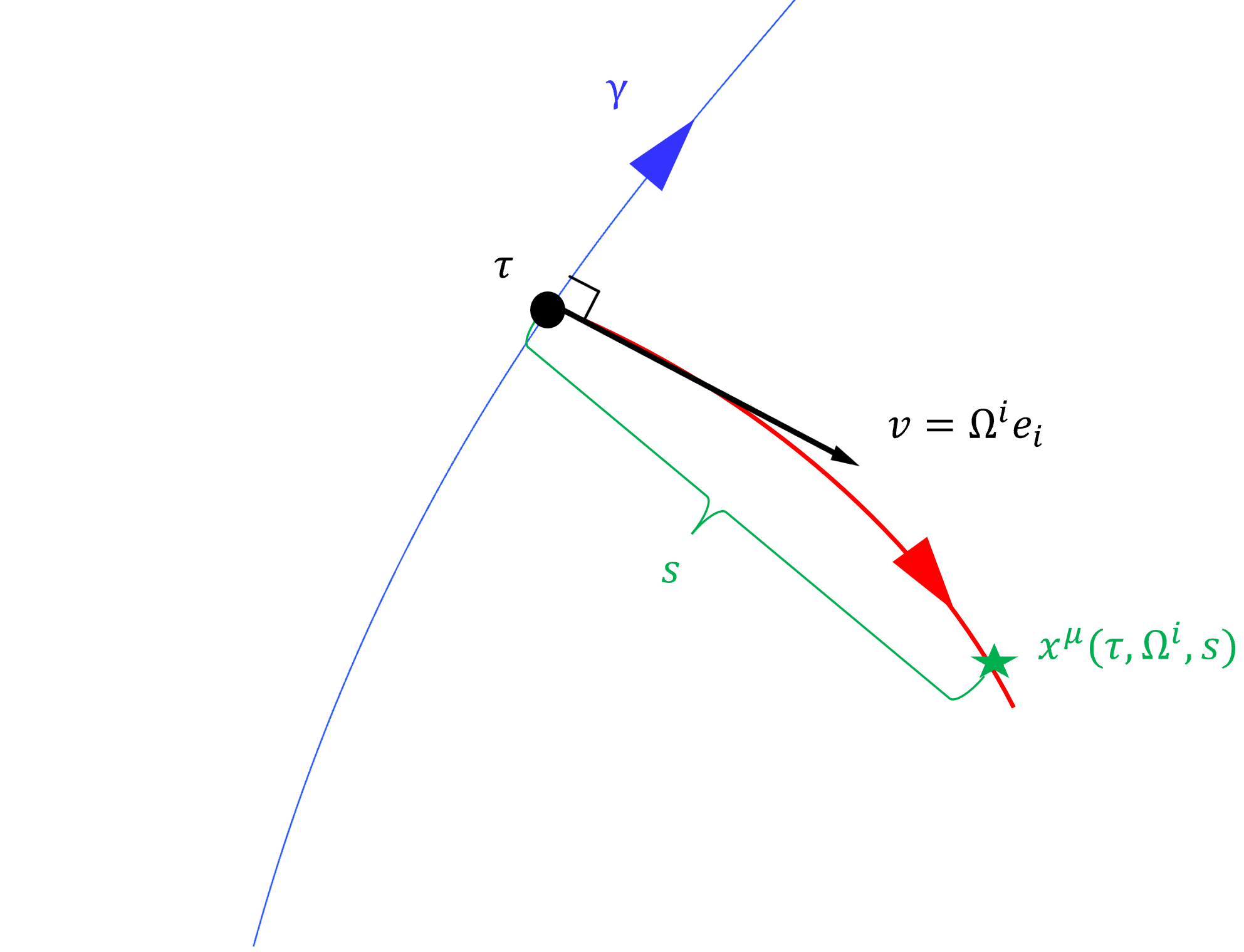}}
\caption{The location of a point on a spacelike geodesic 
as depending on $\tau$, $\Omega^i$ and $s$.}
\label{fig}
\end{figure}


\newpage

Since the background spacetime coordinates at any point on the 
spacelike geodesic must not change under the change of the affine 
parameter $s \rightarrow s/c$ together with the change of the 
initial direction $\Omega^i \rightarrow c\Omega^i$, we conclude 
that
\begin{equation}
x^\mu(\tau,\Omega^i,s) = x^\mu(\tau,c\Omega^i,s/c).            \label{3.7}
\end{equation}
Since $c$ is arbitrary, we can choose $c=s$ so that
\begin{equation}
x^\mu(\tau,\Omega^i,s) = x^\mu(\tau,s\Omega^i,1)               \label{3.8}
\end{equation}
which leads us to conclude that $x^\mu$ is a function of the 
Fermi normal coordinates, $X^0 = \tau$ and $X^i = s\Omega^i$ 
(see Eq. (\ref{3.1})). Eq. (\ref{3.8}) thus indicates a 
one-to-one mapping between the Fermi normal coordinates and 
the background spacetime coordinates,
\begin{equation}
x^\mu = x^\mu(X),                                        \label{3.9}
\end{equation}
in the spacetime region in which the Fermi normal coordinates are 
well defined. This result implies that the tangent vector $v^\mu$ 
in Eq. (\ref{3.2}) can be expressed as
\begin{eqnarray}
v^\mu &=& \left.\frac{dx^\mu}{ds}\right|_{s=0}  \nonumber \\
&=& \left.\frac{\partial x^\mu}{\partial X^a}\frac{dX^a}{ds}\right|_{s=0} 
\nonumber \\
&=& \left.\frac{\partial x^\mu}{\partial X^i}\right|_{s=0}\Omega^i .  
                                                        \label{3.10}
\end{eqnarray}
When combined with Eq. (\ref{3.0}), Eq. (\ref{3.10}) implies that
\begin{equation}
e_i^\mu = \left.\frac{\partial x^\mu}{\partial X^i}\right|_{s=0}. 
                                                        \label{3.11}
\end{equation}
Also, since $e_0$ is, by definition, the tangent vector $u$ of 
the timelike geodesic, 
\begin{equation}
e_0^\mu \equiv u^\mu = \frac{dx^\mu}{d\tau}
\end{equation}
and since $X^0 = \tau$ (see Eq. (\ref{3.1})), we can conclude that
\begin{equation}
e_0^\mu = \left.\frac{\partial x^\mu}{\partial X^0}\right|_{s=0}. 
                                                       \label{3.12}
\end{equation}
Eqs. (\ref{3.11}) and (\ref{3.12}) thus lead us to conclude that 
\begin{equation}
e_a^\mu = \left.\frac{\partial x^\mu}{\partial X^a}\right|_\gamma  
                                                       \label{3.13}
\end{equation}
where the notation $|_\gamma$ indicates that the expression 
must be evaluated on the timelike geodesic. This result enables 
us to obtain any tensorial quantity on the timelike geodesic 
in the Fermi normal coordinate system, provided that we know 
the explicit form of this quantity in the coordinate basis 
and the explicit form of the vierbein. For example, the Riemann 
curvature tensor on the timelike geodesic $\gamma$ in the 
Fermi normal coordinate system is
\begin{eqnarray}
\left. R_{abcd}\right|_\gamma &=& 
\left.\frac{\partial x^\mu}{\partial X^a}\right|_\gamma
\left.\frac{\partial x^\nu}{\partial X^b}\right|_\gamma
\left.\frac{\partial x^\rho}{\partial X^c}\right|_\gamma
\left.\frac{\partial x^\sigma}{\partial X^d}\right|_\gamma
\left. R_{\mu\nu\rho\sigma}\right|_\gamma  \nonumber \\
&=& e_a^\mu e_b^\nu e_c^\rho e_d^\sigma \left. R_{\mu\nu\rho\sigma}
\right|_\gamma                                         \label{3.14}
\end{eqnarray}
with $R_{\mu\nu\rho\sigma}|_\gamma$ being the Riemann curvature 
tensor on $\gamma$ in the coordinates basis. We will use this 
result in the evaluation of the metric tensor in the Fermi 
normal coordinate system below.


\subsection{Geodesic Deviation along the Spacelike Geodesics}

Let us digress a little bit to consider the geodesic deviation 
along the spacelike geodesics, since we will have to use it in 
the evaluation of the metric tensor in the Fermi normal coordinates. 
We first recall that given a point $P$ with an affine parameter 
$\tau$ on a timelike geodesic $\gamma$, there is a one-to-one 
connection between the spacelike geodesics emanating from 
this point and their starting directions $\vec{\Omega} = 
(\Omega^1,\Omega^2,\Omega^3)$, provided that we confine our 
attention to a spacetime region sufficiently close to $\gamma$. 
This implies that the set of all spacelike geodesics emanating 
from each point on $\gamma$ is parametrized by 3 parameters, 
$\Omega^i$ $(i=1,2,3)$, and that we can define the corresponding 
deviation vectors 
\begin{equation}
\xi^\mu_i \equiv \left.\frac{\partial x^\mu}{\partial\Omega^i}
\right|_{\tau,s}                                     \label{3.15}
\end{equation} 
which connects two nearby spacelike geodesics at the same affine 
parameters, $\tau$ and $s$. This is shown in Fig. 3.6 (a). 

On the other hand, we can also consider a set of all spacelike 
geodesics emanating from all points on $\gamma$ (with the affine 
parameter $\tau$) along a fixed direction $\vec{\Omega}$. This family 
of spacelike geodesics is clearly parametrized by $\tau$, with the 
corresponding deviation vector
\begin{equation}
\xi^\mu_\tau \equiv \left.\frac{\partial x^\mu}{\partial\tau}
\right|_{\Omega^i,s}                             \label{3.16}
\end{equation}
as shown in Fig. 3.6 (b). 

\newpage


\begin{figure}[htbp]
\centerline{\includegraphics[scale=.5]{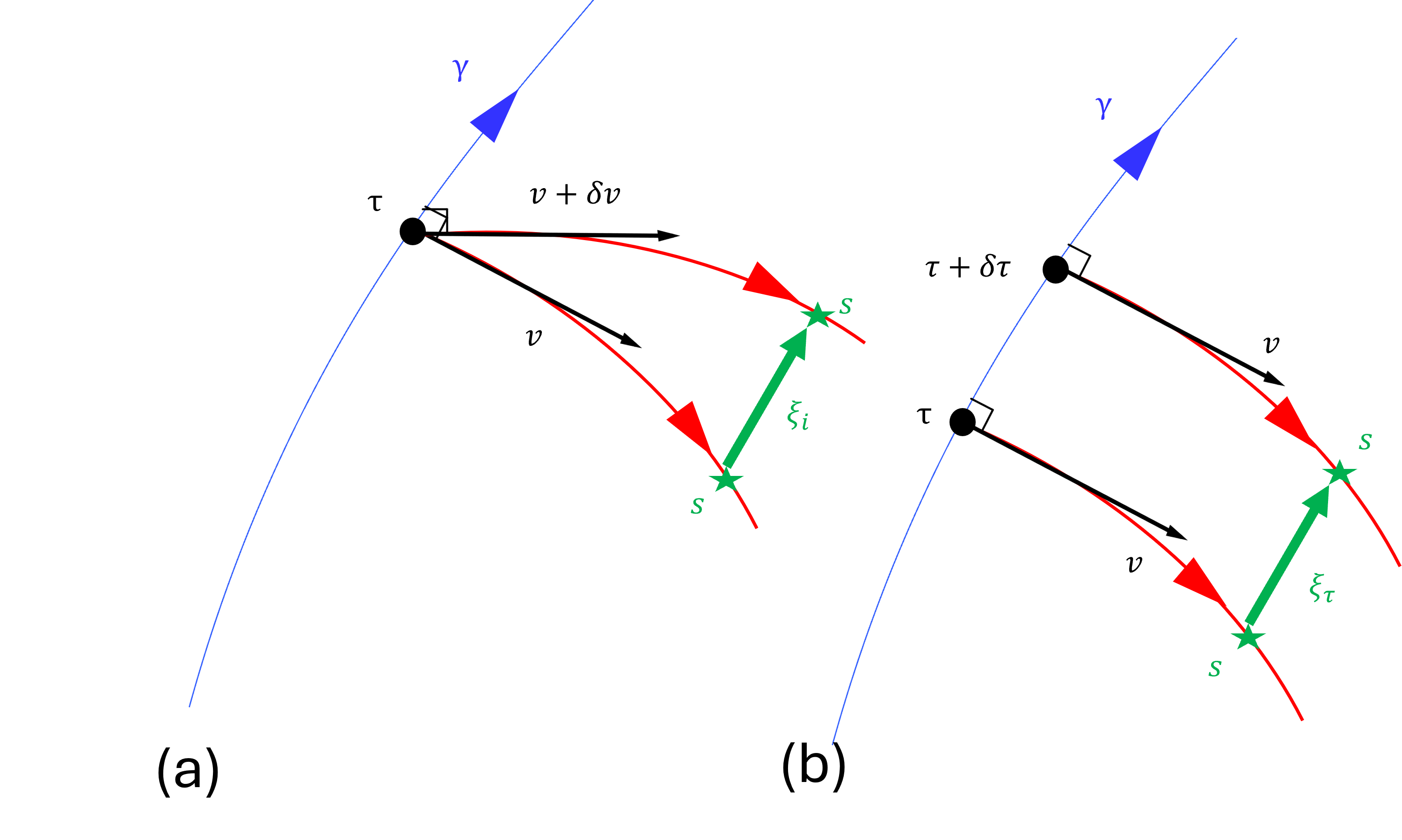}}
\caption{(a) The deviation vector $\xi^\mu_i = 
\partial x^\mu/\partial\Omega^i|_{\tau,s}$ connecting two nearby 
spacelike geodesics, which emanate from the same point on a 
timelike geodesic $\gamma$, at the same affine parameter $s$. 
(b) The deviation vector $\xi^\mu_\tau = \partial x^\mu/\partial 
\tau|_{\Omega^i,s}$ connecting two nearby spacelike geodesics 
emanating from different points on $\gamma$. }
\label{fig}
\end{figure}


\vspace{1cm}

Since both $\xi^\mu_i$ and $\xi^\mu_\tau$ are deviation vectors, 
they satisfy the geodesic deviation equation 
\begin{equation}
\frac{D^2\xi^\mu}{ds^2} = - R^\mu{}_{\nu\rho\sigma}v^\nu\xi^\rho 
v^\sigma ,                                       \label{3.17}
\end{equation}
where $v^\mu \equiv dx^\mu/ds$ is the tangent vector of the 
spacelike geodesic. Let us consider the left-hand side,
\begin{eqnarray}
\frac{D^2\xi^\mu}{ds^2} &=& v^\nu\nabla_\nu (v^\rho\nabla_\rho\xi^\mu) 
\nonumber \\
&=& v^\nu\partial_\nu (v^\rho\nabla_\rho\xi^\mu) + 
v^\nu\Gamma^\mu_{\nu\sigma}(v^\rho\nabla_\rho\xi^\sigma) \nonumber \\
&=& \left\{ \frac{d^2\xi^\mu}{ds^2} + (v^\nu\partial_\nu v^\rho)
\Gamma^\mu_{\rho\sigma}\xi^\sigma + v^\nu v^\rho(\partial_\nu
\Gamma^\mu_{\rho\sigma})\xi^\sigma + v^\rho\Gamma^\mu_{\rho\sigma}
\frac{d\xi^\sigma}{ds} \right\}    \nonumber \\
&& + \left\{ v^\nu\Gamma^\mu_{\nu\rho}\frac{d\xi^\rho}{ds} + 
v^\nu v^\rho\Gamma^\mu_{\nu\sigma}\Gamma^\sigma_{\rho\tau}\xi^\tau 
\right\}                                         \label{3.18}
\end{eqnarray}
where we have used $v^\mu\partial_\mu = d/ds$. Since $v^\mu$ is 
the tangent vector of the geodesic, it satisfies $v^\nu\nabla_\nu 
v^\mu = 0$, which leads to $v^\nu\partial_\nu v^\mu = - v^\nu 
\Gamma^\mu_{\nu\rho}v^\rho$. Using this result in the second 
term on the third line, we obtain
\begin{eqnarray}
\frac{D^2\xi^\mu}{ds^2} &=& \frac{d^2\xi^\mu}{ds^2} + 
2v^\nu\Gamma^\mu_{\nu\rho}\frac{d\xi^\rho}{ds} + 
v^\nu v^\rho (\partial_\nu\Gamma^\mu_{\rho\sigma})\xi^\sigma 
\nonumber \\
&& - v^\nu v^\rho\Gamma^\sigma_{\nu\rho}\Gamma^\mu_{\sigma\tau}
\xi^\tau 
+ v^\nu v^\rho\Gamma^\mu_{\nu\sigma}\Gamma^\sigma_{\rho\tau}\xi^\tau .
                                                \label{3.19}
\end{eqnarray}
Thus
\begin{eqnarray}
R^\mu{}_{\nu\rho\sigma}v^\nu\xi^\rho v^\sigma &=& 
- \frac{d^2\xi^\mu}{ds^2} - 
2v^\nu\Gamma^\mu_{\nu\rho}\frac{d\xi^\rho}{ds} - 
v^\nu v^\rho (\partial_\nu\Gamma^\mu_{\rho\sigma})\xi^\sigma 
\nonumber \\
&& + v^\nu v^\rho\Gamma^\sigma_{\nu\rho}\Gamma^\mu_{\sigma\tau}
\xi^\tau 
- v^\nu v^\rho\Gamma^\mu_{\nu\sigma}\Gamma^\sigma_{\rho\tau}\xi^\tau .
                                                \label{3.20}
\end{eqnarray}

In the Fermi normal coordinate system where $X^0 = \tau$ and 
$X^i = s\Omega^i$, the tangent vector is
\begin{eqnarray}
v^a &=& \frac{dX^a}{ds} \nonumber \\
&=& \Omega^i \delta_i{}^a                       \label{3.21}
\end{eqnarray}
and the deviation vectors are simply
\begin{eqnarray}
\xi^a_i &=& \left.\frac{\partial X^a}{\partial\Omega^i}\right|_{\tau,s} 
\nonumber \\
&=& s\,\delta_i{}^a                               \label{3.22} \\
\xi^a_\tau &=& \left.\frac{\partial X^a}{\partial\tau}\right|_{\Omega^i,s} 
\nonumber \\
&=& \delta_0{}^a .                              \label{3.23}
\end{eqnarray}
Using the above result in Eq. (\ref{3.20}), we obtain 
\begin{equation}
s\!\left( R^a{}_{jik} + \partial_j\Gamma^a_{ki} - \Gamma^b_{jk}\Gamma^a_{bi} 
+ \Gamma^a_{jb}\Gamma^b_{ki} \right)\Omega^j \Omega^k + 2\Omega^j
\Gamma^a_{ji} = 0                               \label{3.24}
\end{equation}
for $\xi^a_i$ and
\begin{equation}
\left( R^a{}_{i0j} + \partial_i\Gamma^a_{j0} - \Gamma^b_{ij}\Gamma^a_{b0} 
+ \Gamma^a_{ib}\Gamma^b_{j0} \right)\Omega^i \Omega^j = 0   \label{3.25}
\end{equation}
for $\xi^a_\tau$. We will use this result in the next subsection.


\subsection{The Metric Tensor in the Fermi Normal Coordinate 
System}

We are now ready to evaluate the metric tensor in the Fermi 
normal coordinate system. The idea is as follows. Since this 
coordinate system is expected to be valid only in a small 
vicinity of the timelike geodesic $\gamma$ (that is, when 
the coordinates $X^i = s\Omega^i$ are expected to be small), 
it is tempting to obtain the Taylor's expansion of 
$g_{ab}(X^0,X^i)$ with respect to the variables $X^i$ of 
the form
\begin{equation}
g_{ab}(\tau,X^i) = \left. g_{ab}\right|_\gamma + \left.\partial_i 
g_{ab}\right|_\gamma X^i + \frac{1}{2}\left.\partial_i
\partial_j g_{ab}\right|_\gamma X^i X^j + \ldots ,      \label{3.26}
\end{equation}
where the notation $|_\gamma$ indicates that the associated 
quantity is evaluated on the timelike geodesic $\gamma$ and 
hence depends on the affine parameter $\tau$. Note that we write 
$\tau$ instead of $X^0$ since this should give us a clearer 
physical picture. (Remember that $X^0 = \tau$ is the proper 
time of the observer moving along the geodesic $\gamma$, 
and so the variable $\tau$ indicates the time at which the 
observer measures the quantity. Using $\tau$ instead of $X^0$ 
thus gives us a better physical feeling.) All we have to 
do is, therefore, to evaluate the spatial derivatives 
of $g_{ab}$ on the geodesic $\gamma$. We do this as follows.
\begin{enumerate}
\item {\bf The metric tensor on the geodesic $\gamma$}: 
Using the transformation law
\begin{equation}
g_{ab} = \frac{\partial x^\mu}{\partial X^a}
\frac{\partial x^\nu}{\partial X^b}g_{\mu\nu},    \label{3.27}
\end{equation}
and Eq. (\ref{3.13}), we simply obtain
\begin{eqnarray}
\left. g_{ab}\right|_\gamma &=& e^\mu_a e^\nu_b \left. 
g_{\mu\nu}\right|_\gamma       \nonumber \\
&=& \eta_{ab}                                     \label{3.28}
\end{eqnarray}
where we have used the orthonormality of the vierbein. 
\item {\bf The first derivative of the metric tensor on 
$\gamma$}: This one is a little more tricky. Consider the 
spacelike geodesic that we used in defining the Fermi normal 
coordinates. In the Fermi normal coordinate system, the 
geodesic equation reads 
\begin{equation}
\frac{d^2 X^a}{ds^2} + \Gamma^a_{bc}\frac{dX^b}{ds}
\frac{dX^c}{ds} = 0. 
\end{equation}
Using $X^0 = \tau$ and $X^i = s\Omega^i$, this equation 
becomes
\begin{equation}
\Gamma^a_{ij}\Omega^i \Omega^j = 0. 
\end{equation}
On the timelike geodesic $\gamma$, $\Gamma^a_{ij}|_\gamma$ 
depends on $\tau$ only and, therefore, must be independent of 
$\Omega^i$. This implies that
\begin{equation}
\left.\Gamma^a_{ij}\right|_\gamma = 0.            \label{3.29}
\end{equation}
We next find the other components of the connection on 
$\gamma$. Consider the parallel-transportation equation 
defining the vierbein, 
\begin{equation}
u^\nu\nabla_\nu e_a^\mu = \frac{de_a^\mu}{d\tau} + 
u^\nu\Gamma^\mu_{\nu\rho}e_a^\rho = 0,            \label{3.30}
\end{equation}
along the timelike geodesic. In the Fermi normal coordinate 
system, Eq. (\ref{3.13}) tells us that $e_a^b = 
(\partial X^b/\partial X^a)|_\gamma = \delta_a{}^b$ and $u^a 
= (\partial X^a/\partial\tau)|_\gamma = \delta_0{}^a$. Using 
these in Eq. (\ref{3.30}), we find
\begin{equation}
\left.\Gamma^a_{0b}\right|_\gamma = 0.            \label{3.31}
\end{equation}
From Eqs. (\ref{3.29}) and (\ref{3.31}), we conclude that 
all components of the connection in the Fermi normal coordinate 
system on $\gamma$ are zero,
\begin{equation}
\left.\Gamma^a_{bc}\right|_\gamma = 0.           \label{3.32}
\end{equation}
To find $\partial_a g_{bc}|_\gamma$, we recall that the 
metric tensor satisfies the metric compatibility condition, 
$\nabla_\mu g_{\nu\rho} = 0$. In the Fermi normal coordinate 
system, this condition reads
\begin{equation}
\partial_a g_{bc} - \Gamma^d_{ab}g_{dc} - \Gamma^d_{ac}g_{bd} 
= 0.                                             \label{3.33}
\end{equation}
Evaluating this equation on $\gamma$ and using Eq. (\ref{3.32}), 
we finally obtain
\begin{equation}
\left.\partial_a g_{bc}\right|_\gamma = 0.       \label{3.34}
\end{equation}
From Eqs. (\ref{3.28}) and (\ref{3.34}), we see that the Fermi 
normal coordinate system provides an example of the local flatness 
theorem \cite{Schutz}.  
\item {\bf The second derivative of the metric tensor on 
$\gamma$}: The idea is simple. Differentiating Eq. (\ref{3.33}) 
and using Eqs. (\ref{3.28}) and (\ref{3.32}), we find
\begin{equation}
\left.\partial_a\partial_b g_{cd}\right|_\gamma = 
\left.(\partial_a\Gamma^e_{bc})\right|_\gamma\eta_{ed} + 
\left.(\partial_a\Gamma^e_{bd})\right|_\gamma\eta_{ce},   \label{3.35}
\end{equation}
which implies that finding $\partial_a\partial_b g_{cd}|_\gamma$ 
is equivalent to finding $(\partial_a\Gamma^d_{bc})|_\gamma$. 
Let us find $(\partial_a\Gamma^d_{bc})|_\gamma$ step by step 
as follows.
\begin{itemize}
\item Recall that $\Gamma^a_{bc}|_\gamma = 0$ (see Eq. (\ref{3.32})). 
In other words, $\Gamma^a_{bc}$ is constant along $\gamma$, 
which implies 
\begin{equation}
\left.\partial_0\Gamma^a_{bc}\right|_\gamma = 0.    \label{3.36}
\end{equation}
\item From the definition of the Riemann curvature tensor,
\begin{equation}
R^a{}_{bcd} = \partial_c\Gamma^a_{db} - \partial_d\Gamma^a_{cb} 
+ \Gamma^a_{ce}\Gamma^e_{db} - \Gamma^a_{de}\Gamma^e_{cb}.   \label{3.37}
\end{equation}
Using Eqs. (\ref{3.32}) and (\ref{3.36}) and setting $d=0$ 
in Eq. (\ref{3.37}), we find
\begin{equation}
\left.\partial_c\Gamma^a{}_{0b}\right|_\gamma = \left. R^a{}_{bc0}
\right|_\gamma .                                   \label{3.38}
\end{equation}
\item Recall Eq. (\ref{3.24}),
\begin{equation}
s\!\left( R^a{}_{jik} + \partial_j\Gamma^a_{ki} - 
\Gamma^b_{jk}\Gamma^a_{bi} + \Gamma^a_{jb}\Gamma^b_{ki} \right)
\Omega^j\Omega^k + 2\Omega^j\Gamma^a_{ji} = 0.     \label{3.39}
\end{equation}
Expand $\Gamma^a_{ji}$ in the last term as a power series 
in $s$ as
\begin{eqnarray}
\Gamma^a_{ji} &=& \left.\Gamma^a_{ji}\right|_\gamma + 
s\frac{dX^k}{ds}\left.\partial_k\Gamma^a_{ji}\right|_\gamma 
+ \ldots    \nonumber \\
&=& \left.\Gamma^a_{ji}\right|_\gamma + 
s\Omega^k\left.\partial_k\Gamma^a_{ji}\right|_\gamma + \ldots ,
\end{eqnarray}
and substitute the result into Eq. (\ref{3.39}). To the first 
order in $s$, Eq. (\ref{3.39}) becomes
\begin{equation}
\left. (R^a{}_{jik} + 3\partial_j\Gamma^a_{ki})\right|_\gamma
\Omega^j\Omega^k = 0,                             \label{3.40}
\end{equation}
where we have used the fact that the connection vanishes 
on $\gamma$. Using the fact that $\Omega^j\Omega^k$ is 
symmetric in $j$ and $k$, we find
\begin{equation}
\left.\left(\partial_j\Gamma^a_{ki} + \partial_k\Gamma^a_{ji}
\right)\right|_\gamma = -\frac{1}{3}\left.\left(
R^a{}_{jik} + R^a{}_{kij}\right)\right|_\gamma .  \label{3.41}
\end{equation}
On the other hand, Eq. (\ref{3.37}) implies that
\begin{equation}
\left.\left(\partial_j\Gamma^a_{ki} - \partial_k\Gamma^a_{ji}
\right)\right|_\gamma = \left. R^a{}_{ijk}\right|_\gamma 
                                                 \label{3.42}
\end{equation}
which is antisymmetric in $j$ and $k$. Using Eqs. (\ref{3.41}) 
and (\ref{3.42}), we arrive at
\begin{equation}
\left.\partial_i\Gamma^a_{jk}\right|_\gamma = -\frac{1}{3}
\left.\left(R^a{}_{jki} + R^a{}_{kji}\right)\right|_\gamma . 
                                                 \label{3.43}
\end{equation}
\end{itemize}
Using the results (\ref{3.36}), (\ref{3.38}) and (\ref{3.43}) 
in Eq. (\ref{3.35}), we finally obtain
\begin{eqnarray}
\left.\partial_i\partial_j g_{00}\right|_\gamma &=& 
-2\left. R_{0i0j}\right|_\gamma                 \label{3.44} \\
\left.\partial_i\partial_j g_{0k}\right|_\gamma &=&
-\frac{2}{3}\left.\left(R_{0ikj} + R_{0jki}\right)\right|_\gamma 
                                                \label{3.45} \\
\left.\partial_i\partial_j g_{kl}\right|_\gamma &=&
-\frac{1}{3}\left.\left(R_{kilj} + R_{kjli}\right)\right|_\gamma ,
                                                \label{3.46}
\end{eqnarray}
where we have lowered the index of the Riemann tensor using 
the metric $\eta_{ab}$ on $\gamma$. This says that once we 
know all components of the Riemann tensor in the Fermi normal 
coordinate system on the timelike geodesic $\gamma$, $\partial_i
\partial_j g_{ab}|_\gamma$ will follow immediately. In practical 
calculations, one begins by calculating the Riemann tensor in 
the coordinate basis, and then converts the result to the one 
in the Fermi normal coordinate system on $\gamma$ by using 
Eq. (\ref{3.14}). An example of this kind of calculation 
will be shown later on in this chapter. 
\end{enumerate}

Using Eqs. (\ref{3.28}), (\ref{3.34}), (\ref{3.44}), (\ref{3.45}) 
and (\ref{3.46}) in Eq. (\ref{3.26}), we finally obtain the 
metric tensor in the Fermi normal coordinate system,
\begin{eqnarray}
g_{00}(\tau,X^i) &=& 1 - R_{0i0j}(\tau)X^iX^j + {\cal O}(X^3) 
                                               \label{3.47} \\
g_{0i}(\tau,X^i) &=& -\frac{2}{3}R_{0jik}(\tau)X^jX^k + 
{\cal O}(X^3)                                  \label{3.48} \\
g_{ij}(\tau,X^i) &=& -\delta_{ij} - \frac{1}{3}R_{ikjl}(\tau)
X^kX^l + {\cal O}(X^3),                        \label{3.49}
\end{eqnarray}
where we have replaced the notation $|_\gamma$ by the functional 
dependence on $\tau$ to indicate the dependence on the observer's 
proper time. This concludes our discussion of the Fermi normal 
coordinates. 


\section{Motion of a Free Particle in the Fermi Normal Coordinate 
System}

Having obtained all the machinery in the previous section, we 
are now ready to discuss the motion of a free particle as seen 
by an observer moving along the timelike geodesic $\gamma$. As 
mentioned before, this observer uses the Fermi normal coordinates 
to describe things that he observes. Therefore, to find out what 
this observer sees, we consider a geodesic equation in the Fermi 
normal coordinate system describing the motion of a free particle 
with the proper time $\tilde{s}$,
\begin{equation}
\frac{d^2X^a}{d\tilde{s}^2} + \Gamma^a_{bv}\frac{dX^b}{d\tilde{s}}
\frac{dX^c}{d\tilde{s}} = 0.                               \label{3.50}
\end{equation}
Since the observer never knows the value of $\tilde{s}$ and all 
he knows is that $\tilde{s}$ must be an increasing function of 
his proper time $\tau$ (which is $X^0$), it is tempting to change 
the variable in Eq. (\ref{3.50}) from $\tilde{s}$ to $\tau$. To 
do this, we first define
\begin{equation}
\Sigma \equiv \frac{d\tau}{d\tilde{s}}.                       \label{3.51}
\end{equation}
From the metric on the particle's path,
\begin{eqnarray}
d\tilde{s}^2 &=& g_{ab}dX^adX^b   \nonumber \\
&=& g_{00}d\tau^2 + 2g_{0i}d\tau dX^i + g_{ij}dX^idX^j ,  \label{3.52}
\end{eqnarray}
we obtain
\begin{eqnarray}
\frac{1}{\Sigma^2} &=& \left(\frac{d\tilde{s}}{d\tau}\right)^2  \nonumber \\
&=& g_{00} + 2g_{0i}V^i + g_{ij}V^iV^j ,                  \label{3.53}
\end{eqnarray}
where $V^i \equiv dX^i/d\tau$ is the particle's velocity as 
measured by the observer. Changing the variable from $\tilde{s}$ 
to $\tau$ in Eq. (\ref{3.50}), we get 
\begin{equation}
\frac{d^2X^a}{d\tau^2} + \frac{1}{\Sigma}\frac{d\Sigma}{d\tau}
\frac{dX^a}{d\tau} + \Gamma^a_{bc}\frac{dX^b}{d\tau}\frac{dX^c}{d\tau} 
= 0,                                                      \label{3.54}
\end{equation}
which clearly contains only the quantities that can be measured 
by the observer. Eq. (\ref{3.54}) thus describes the law of motion 
of a free particle as seen by the observer moving along a timelike 
geodesic. Setting $a=0$ in Eq. (\ref{3.54}) and using $X^0 = \tau$, 
we get 
\begin{equation}
\frac{1}{\Sigma}\frac{d\Sigma}{d\tau} = - 
\Gamma^0_{bc}\frac{dX^b}{d\tau}\frac{dX^c}{d\tau}.        \label{3.55}
\end{equation}
Setting $a=i$ in Eq. (\ref{3.54}) and substituting the result 
(\ref{3.55}), we obtain
\begin{equation}
\frac{d^2X^i}{d\tau^2} + \left(\Gamma^i_{ab} - \Gamma^0_{ab}V^i\right)
\frac{dX^a}{d\tau}\frac{dX^b}{d\tau} = 0                  \label{3.56}
\end{equation}
which is precisely the equation of motion of the particle as 
described by the observer. To see the precise form of this 
equation of motion, we need to evaluate the Christoffel 
connection in the Fermi normal coordinate system. Using the 
metric tensor in Eqs. (\ref{3.47})--(\ref{3.49}) and noting 
that it is of the form of the metric in the linearized gravity 
($g_{\mu\nu} = \eta_{\mu\nu} + h_{\mu\nu}$ with the inverse 
$g^{\mu\nu} = \eta^{\mu\nu} - \eta^{\mu\rho}\eta^{\nu\sigma}
h_{\rho\sigma}$), we immediately obtain the inverse metric 
tensor in the Fermi normal coordinate system,
\begin{eqnarray}
g^{00}(\tau,X^i) &=& 1 + R_{0i0j}(\tau)X^iX^j + {\cal O}(X^3) 
                                                         \label{3.57} \\
g^{0i}(\tau,X^i) &=& -\frac{2}{3}R_{0jik}(\tau)X^j X^k + {\cal O}(X^3) 
                                                         \label{3.58} \\
g^{ij}(\tau,X^i) &=& -\delta_{ij} + \frac{1}{3}R_{ikjl}(\tau)X^k X^l 
+ {\cal O}(X^3).                                         \label{3.59}
\end{eqnarray}
With this result, we are ready to calculate the Christoffel 
connection,
\begin{equation}
\Gamma^a_{bc} = \frac{1}{2}g^{ad}(\partial_b g_{cd} + \partial_c g_{bd} 
- \partial_d g_{bc}).                                    \label{3.60}
\end{equation}
But before we do this, it is important to recall a familiar 
result that the metric perturbation $h_{\mu\nu}$ plays the 
role of the gravitational potential in the non-relativistic 
limit \cite{Schutz}. As we have expanded the metric tensor 
only up to the second order in $X^i$, it is clear that the 
form of the corresponding gravitational force is exact only 
up to the linear order in $X^i$. Eq. (\ref{3.56}) thus implies 
that we need to calculate the connection only up to the 
linear order in $X^i$ so as to make the linearized formalism 
consistent. With this in mind, we calculate the Christoffel 
connection up to the first order in $X^i$, with the result:
\begin{eqnarray}
\Gamma^0_{00} &=& 0   \\
\Gamma^0_{i0} &=& -R_{0i0j}(\tau)X^j   \\
\Gamma^0_{ij} &=& -\frac{1}{3}(R_{0ijk}(\tau) + R_{0jik}(\tau))X^k  \\
\Gamma^i_{00} &=& -R_{0i0j}(\tau)X^j   \\
\Gamma^i_{j0} &=& R_{0kij}(\tau)X^k    \\
\Gamma^i_{jk} &=& \frac{1}{3}(R_{ijkl}(\tau) + R_{ikjl}(\tau))X^l. 
\end{eqnarray}
Substitute the above result into Eq. (\ref{3.56}) and keep 
the terms up to the first order in $V^i$ (which is indeed 
the velocity of the particle divided by the speed of light), 
we obtain
\begin{equation}
\frac{d^2X^i}{d\tau^2} = R_{0i0j}(\tau)X^j + 2R_{ijk0}(\tau)X^k V^j. 
                                                      \label{3.61}
\end{equation}
To rewrite the above result in a more familiar form, we 
define the gravitoelectric field $\vec{E}_G$ by 
\begin{equation}
(E_G)_i(\tau,\vec{X}) \equiv R_{0i0j}(\tau)X^j        \label{3.62}
\end{equation}
and the gravitomagnetic field $\vec{B}_G$ by 
\begin{equation}
(B_G)_i(\tau,\vec{X}) \equiv -\frac{1}{2}\epsilon_{ijk}R_{jk0l}
(\tau)X^l.                                            \label{3.63}
\end{equation}
Multiply both sides of Eq. (\ref{3.61}) by the particle's 
rest mass $m$, the equation of motion (\ref{3.56}) finally 
takes the form
\begin{equation}
m\frac{d^2\vec{X}}{d\tau^2} = q_E\vec{E}_G + 
q_B\,\vec{V}\times \vec{B}_G ,                        \label{3.64}
\end{equation}
where $q_E \equiv m$ and $q_B \equiv 2m$. This tells us 
that the observer moving along a timelike geodesic will 
see another free-particle nearby him as being subject to 
the gravitoelectric and gravitomagnetic forces. In other 
words, a free-falling observer, who is unaware of the 
phenomenon of gravitoelectromagnetism, will not believe 
that this particle is indeed a free particle. In the 
next section, we will perform a detailed calculation 
to find the gravitoelectric and gravitomagnetic fields 
in linearized Rastall gravity.


\section{Gravitoelectromagnetism in Linearized Rastall 
Gravity}

Having obtained the expression for the gravitoelectric 
and gravitomagnetic fields in the last section, let us 
now calculate these fields in the linearized Rastall 
gravity. We will limit ourselves to the simplest case 
of the circular motion around a non-rotating mass, 
described by the metric
\begin{equation}
ds^2 = (1-\alpha\Phi(r))dt^2 - (1+\beta\Phi(r))dr^2 - 
r^2d\theta^2 - r^2\sin^2\theta d\phi^2 ,        \label{3.65}
\end{equation}
where $\Phi(r) = GM/r$ and
\begin{eqnarray}
\alpha &\equiv & 2\left(\frac{1-6\lambda}{1-4\lambda}\right) 
                                                \label{3.66} \\
\beta &\equiv & 2\left(\frac{1-2\lambda}{1-4\lambda}\right), 
                                                \label{3.67}
\end{eqnarray}
derived in Chapter 2. The reason why we limit ourselves 
to the circular orbit is that, from our experience with 
the Schwarzschild metric in General Relativity, it is 
the only possible closed orbit, whose form of the 
worldline can exactly be found.

Our first step is to find the Riemann curvature tensor 
in the coordinate basis. Let $f(r)\equiv (1-\alpha\Phi(r))$ 
and $g(r)\equiv (1+\beta\Phi(r))$, we find the following 
non-zero components of the Christoffel connection of the 
metric (\ref{3.65}),
\begin{eqnarray}
\Gamma^t_{rt} = \Gamma^t_{tr} &=& \frac{1}{2f}\frac{df}{dr} \\
\Gamma^r_{tt} &=& \frac{1}{2g}\frac{df}{dr}  \\
\Gamma^r_{rr} &=& \frac{1}{2g}\frac{dg}{dr}  \\
\Gamma^r_{\theta\theta} &=& -\frac{r}{g}   \\
\Gamma^r_{\phi\phi} &=& - \frac{r\sin^2\theta}{g}  \\
\Gamma^\theta_{r\theta} = \Gamma^\theta_{\theta r} &=& 
\frac{1}{r} \\
\Gamma^\theta_{\phi\phi} &=& -\sin\theta\cos\theta  \\
\Gamma^\phi_{r\phi} = \Gamma^\phi_{\phi r} &=& 
\frac{1}{r}  \\
\Gamma^\phi_{\theta\phi} = \Gamma^\phi_{\phi\theta} &=& 
\cot\theta ,
\end{eqnarray}
from which we can calculate the Riemann curvature tensor
\begin{equation}
R^\mu{}_{\nu\rho\sigma} = \partial_\rho\Gamma^\mu_{\sigma\nu} 
- \partial_\sigma\Gamma^\mu_{\rho\nu} + \Gamma^\mu_{\rho\tau}
\Gamma^\tau_{\sigma\nu} - \Gamma^\mu_{\sigma\tau}
\Gamma^\tau_{\rho\nu}. 
\end{equation}
The non-zero components of $R_{\mu\nu\rho\sigma} = g_{\mu\tau}
R^\tau{}_{\nu\rho\sigma}$ are found to be
\begin{eqnarray}
R_{trtr} &=& -\frac{1}{2}\frac{d^2f}{dr^2} + \frac{1}{4g} 
\frac{df}{dr}\frac{dg}{dr} + \frac{1}{4f}\left(
\frac{df}{dr}\right)^2                              \label{3.68} \\
R_{t\theta t\theta} &=& -\frac{r}{2g}\frac{df}{dr}  \label{3.69} \\
R_{t\phi t\phi} &=& -\frac{r\sin^2\theta}{2g}
\frac{df}{dr}                                       \label{3.70} \\
R_{r\theta r\theta} &=& -\frac{r}{2g}\frac{dg}{dr}  \label{3.71} \\
R_{r\phi r\phi} &=& -\frac{r\sin^2\theta}{2g}
\frac{dg}{dr}                                       \label{3.72} \\
R_{\theta\phi\theta\phi} &=& -r^2\sin^2\theta
\left(\frac{g-1}{g}\right),                         \label{3.73}
\end{eqnarray}
together with other components related to the above by 
index permutation, i.e., $R_{\mu\nu\rho\sigma} = 
- R_{\nu\mu\rho\sigma} = - R_{\mu\nu\sigma\rho} = 
R_{\rho\sigma\mu\nu}$. 

Our next step is to find a closed orbit of the metric 
(\ref{3.65}). This is done by solving the geodesic 
equation for a timelike geodesic with the proper time 
$\tau$. Since $g_{\mu\nu}$ is independent of $t$ 
and $\phi$, we know from \cite{Schutz} that the 
geodesic equation implies that the quantities 
$u_t \equiv g_{t\mu}dx^\mu/d\tau$ and $u_\phi \equiv 
g_{\phi\mu}dx^\mu/d\tau$ are constant, and that 
the spherical symmetry of the metric implies that 
we can choose the orbit to lie in a two-dimensional 
plane $\theta = \pi/2$. Thus setting $u_t = E$, 
$u_\phi = -L$ and $\theta = \pi/2$, we find
\begin{equation}
\frac{dt}{d\tau} \equiv u^t = g^{tt}u_t = 
\frac{E}{(1-\alpha\Phi(r))}                       \label{3.74}
\end{equation}
and
\begin{equation}
\frac{d\phi}{d\tau} \equiv u^\phi = g^{\phi\phi}
u_\phi = \frac{L}{r^2}.                           \label{3.75}
\end{equation}
By using Eqs. (\ref{3.74}) and (\ref{3.75}) and 
setting $\theta = \pi/2$ in the worldline
\begin{equation}
d\tau^2 = (1-\alpha\Phi(r))dt^2 - (1+\beta\Phi(r))dr^2 - 
r^2d\theta^2 - r^2\sin^2\theta d\phi^2, 
\end{equation}
we obtain
\begin{eqnarray}
1 &=& (1-\alpha\Phi(r))(u^t)^2 - (1+\beta\Phi(r))\left(
\frac{dr}{d\tau}\right)^2 - r^2(u^\phi)^2    \nonumber \\
&=& \frac{E^2}{(1-\alpha\Phi(r))} - (1+\beta\Phi(r))
\left(\frac{dr}{d\tau}\right)^2 - \frac{L^2}{r^2},  \label{3.76}
\end{eqnarray}
which can be written as
\begin{equation}
\left(\frac{dr}{d\tau}\right)^2 + V(r) = 0,        \label{3.77}
\end{equation}
where
\begin{equation}
V(r) \equiv \frac{1}{(1+\beta\Phi(r))}\left(1+\frac{L^2}{r^2}
\right) - \frac{E^2}{(1-\alpha\Phi(r))(1+\beta\Phi(r))}. 
                                                   \label{3.78}
\end{equation}
Eqs. (\ref{3.76})--(\ref{3.78}) together with the condition 
$dr/d\tau = 0$ for the circular orbit lead us to conclude 
that the constant radius $r_0$ of the circular orbit must 
satisfy the condition $V(r_0) = 0$, which enables us to 
express $E$ in terms of $L$ and $r_0$ as 
\begin{equation}
E^2 = (1-\alpha\Phi(r_0))\left(1+\frac{L^2}{r_0^2}\right). 
                                                   \label{3.79}
\end{equation}
To find the radius $r_0$, we need to impose one more 
condition. Observe that the left-hand side of Eq. (\ref{3.77}) 
takes the form of the sum of the kinetic energy and 
the potential energy in classical mechanics. This 
implies that the radius $r_0$ that we look for corresponds 
to a local minimum of $V(r)$ satisfying $dV(r)/dr|_{r=r_0} 
= 0$, which has a physical meaning that the radial force 
is zero. The second conditon that we need is thus 
$dV(r)/dr|_{r=r_0} = 0$. By using Eq. (\ref{3.79}) and 
$\Phi(r_0) = GM/r_0$ in $dV(r)/dr|_{r=r_0} = 0$, we 
arrive at 
\begin{equation}
\frac{2L^2}{r_0^3}\left(1-\frac{\alpha GM}{r_0}\right) - 
\alpha\left(1+\frac{L^2}{r_0^2}\right)\frac{GM}{r_0^2} 
= 0, 
\end{equation}
which can be solved for $r_0$ to obtain
\begin{equation}
r_0^{(\pm )} = L \left( \frac{1 \pm \sqrt{1-3(\alpha GM/L)^2}}
{(\alpha GM/L)} \right).                       \label{3.79.1}
\end{equation}
To determine whether we should choose the plus or the 
minus sign in Eq. (\ref{3.79.1}), let us consider a 
circular orbit in the non-relativistic limit. Let 
$v = r_0 d\phi/dt$ be the orbital speed along the 
circular orbit, then the radius $r_0$ of the circular 
orbit is related to $v$ by 
\begin{equation}
\frac{GM}{r_0^2} = \frac{v^2}{r_0},            \label{3.79.2}
\end{equation}
which gives $GM/r_0 = v^2$. By restoring the speed of 
light $c$ to Eq. (\ref{3.75}) (which changes $\tau\to c\tau$) 
and recalling that $dt \approx d\tau$ in the non-relativistic 
limit, Eq. (\ref{3.75}) gives  
\begin{eqnarray}
L &=& \frac{r_0}{c}\left(r_0\frac{d\phi}{d\tau}\right) 
\nonumber \\ 
&\approx & r_0\left(\frac{v}{c}\right).        \label{3.79.3}
\end{eqnarray}
Also, restoring $c$ to the metric tensor changes $GM$ to 
$GM/c^2$. Eqs. (\ref{3.79.2}) and (\ref{3.79.3}) thus 
lead to 
\begin{eqnarray}
\frac{GM}{c^2L} &=& \frac{GM/c^2r_0}{L/r_0}  \nonumber \\
&\approx & \frac{v^2/c^2}{v/c}  \nonumber \\
&=& \frac{v}{c}.                               \label{3.79.4} 
\end{eqnarray}
Since $\alpha = 2$ in Einstein gravity ($\lambda = 0$) 
and $v/c \ll 1$ in the non-relativistic limit, Eqs. 
(\ref{3.79.1}), (\ref{3.79.3}) and (\ref{3.79.4}) imply 
that the non-relativistic limit is consistent only if 
we choose the plus sign in Eq. (\ref{3.79.1}). Thus the 
radius of the circular orbit is
\begin{equation}
r_0 = L \left( \frac{1 + \sqrt{1-3(\alpha GM/L)^2}}
{(\alpha GM/L)} \right),                       \label{3.80}
\end{equation}
which will be substituted into Eq. (\ref{3.79}) to 
find $E$. 

We now summarize our result for the circular orbit 
as follows. For a given value of $L$, let
\begin{equation}
a \equiv \frac{\alpha GM}{L}.                  \label{3.81}
\end{equation}
We calculate the radius of the circular orbit from 
\begin{equation}
r_0 = L\left(\frac{1+\sqrt{1-3a^2}}{a}\right), \label{3.82}
\end{equation}
and substitute the result into Eq. (\ref{3.79}) to 
obtain
\begin{eqnarray}
E &=& \sqrt{\left(1-\frac{\alpha GM}{r_0}\right)
\left(1+\frac{L^2}{r_0^2}\right)}  \nonumber \\
&=& \sqrt{\left(1-a\left(\frac{L}{r_0}\right)\right)
\left(1+\left(\frac{L}{r_0}\right)^2\right)}.  \label{3.83}
\end{eqnarray}
From Eqs. (\ref{3.74}) and (\ref{3.75}), the tangent 
vector to the corresponding timelike geodesic in the 
coordinate basis is
\begin{eqnarray}
u &=& u^t\,\frac{\partial}{\partial t} + 
u^\phi\,\frac{\partial}{\partial\phi}    \nonumber \\
&=& \frac{E}{(1-(\alpha GM/r_0))}\,\frac{\partial}{\partial t} 
+ \frac{L}{r_0^2}\,\frac{\partial}{\partial\phi}. 
                                               \label{3.84}
\end{eqnarray}

To find the vierbein along this geodesic, it is 
appropriate to change the spatial coordinates from 
spherical coordinates $(r,\theta,\phi)$ to cylindrical 
coordinates $(\rho,\phi,z)$ as the orbit is in the 
$\theta = \pi/2$ (or $z=0$) plane. The metric now 
takes the form
\begin{eqnarray}
ds^2 &=& (1-\alpha\Phi)dt^2 - 
\left(1+\frac{\beta\Phi\rho^2}{(\rho^2+z^2)}\right)d\rho^2 
- \left(1+\frac{\beta\Phi z^2}{(\rho^2+z^2)}\right)dz^2 
\nonumber \\
&& - \frac{2\beta\Phi\rho z}{(\rho^2+z^2)}d\rho dz 
- \rho^2d\phi^2 ,                             \label{3.85}
\end{eqnarray}
where $\Phi = GM/\sqrt{\rho^2 + z^2}$. 
As the basis vector $\partial/\partial\phi$ is the 
same in both coordinate systems, the tangent vector 
of the timelike geodesic remains of the form (\ref{3.84}). 
Thus the zero-component of the vierbein, $e_0 = u$, is
\begin{equation}
e_0 = \frac{E}{(1-(\alpha GM/r_0))}\,\frac{\partial}
{\partial t} + \frac{L}{r_0^2}\,\frac{\partial}{\partial\phi}.
                                              \label{3.86}
\end{equation}
To find the other components of the vierbein, we 
first recall that a timelike vector $T = a\,
\partial/\partial t + b\,\partial/\partial x$ 
(with $a > b$) and a spacelike vector 
$V = b\,\partial/\partial t + 
a\,\partial/\partial x$ are orthogonal in the 
Minkowski space. This motivates us to construct 
a spacelike vector pointing along the 
$\phi$-direction and orthogonal to $u$ by swapping 
the components of $u$ in Eq. (\ref{3.84}) and 
then modifying the result by adding some appropriate 
factors from the metric tensor. In this way, we 
obtain the vierbein $e_2$ satisfying $e_0\!\cdot\!e_2
= 0$ and $e_2\!\cdot\!e_2 = -1$ on the $z=0$ plane 
and at $\rho = r_0$ as 
\begin{eqnarray}
e_2 &=& \frac{r_0 u^\phi}{\sqrt{1-(\alpha GM/r_0)}}\,
\frac{\partial}{\partial t} + \frac{\sqrt{1-(\alpha GM/r_0)}
u^t}{r_0}\,\frac{\partial}{\partial\phi}  \nonumber \\
&=& \frac{1}{\sqrt{1-(\alpha GM/r_0)}}\left(\frac{L}{r_0}\,
\frac{\partial}{\partial t} + \frac{E}{r_0}\,
\frac{\partial}{\partial\phi}\right).         \label{3.87}
\end{eqnarray}
Note that we call it $e_2$ since it has a spatial 
component pointing along $\phi\equiv x^2$ direction. 
It is easy to see that the other two spacelike 
components of the vierbein, which we call $e_1$ 
and $e_3$, must point along $\rho$ and $z$ directions 
since they are automatically orthogonal to both $e_0$ 
and $e_2$. By demanding that $e_1\!\cdot\!e_1 = 
-1 = e_3\!\cdot\!e_3$ on the $z=0$ plane and at 
$\rho = r_0$, we find
\begin{eqnarray}
e_1 &=& \frac{1}{\sqrt{1+(\beta GM/r_0)}}\,
\frac{\partial}{\partial\rho}                  \label{3.88} \\
e_3 &=& \frac{\partial}{\partial z}.           \label{3.89}
\end{eqnarray}
Eqs. (\ref{3.86})--(\ref{3.89}) constitute the vierbein 
along the spacelike geodesic corresponding to the 
circular orbit, satisfying $e_a\!\cdot\!e_b = \eta_{ab}$. 
We now change the basis to the coordinate basis in 
spherical coordinates $(r,\theta,\phi)$. Using 
$r = \sqrt{\rho^2 + z^2}$ and $\tan\theta = \rho /z$, 
we find
\begin{eqnarray}
\frac{\partial}{\partial\rho} &=& \sin\theta\,
\frac{\partial}{\partial r} + \frac{\cos\theta}{r}\,
\frac{\partial}{\partial\theta}               \label{3.90} \\
\frac{\partial}{\partial z} &=& \cos\theta\,
\frac{\partial}{\partial r} - \frac{\sin\theta}{r}\,
\frac{\partial}{\partial\theta}.               \label{3.91}
\end{eqnarray}
Along the geodesic, $\theta = \pi/2$ and $r=r_0$, so that 
we can substitute $\partial/\partial\rho = \partial/\partial r$ 
and $\partial/\partial z = -(1/r_0)\partial/\partial\theta$ 
into Eqs. (\ref{3.88})--(\ref{3.89}) to obtain $e_1$ and 
$e_3$ in the spherical coordinates. We thus obtain the 
vierbein in the spherical coordinate basis,
\begin{eqnarray}
e_0 &=& \frac{E}{(1-(\alpha GM/r_0))}\,\frac{\partial}
{\partial t} + \frac{L}{r_0^2}\,\frac{\partial}{\partial\phi}
                                               \label{3.92} \\
e_1 &=& \frac{1}{\sqrt{1+(\beta GM/r_0)}}\,
\frac{\partial}{\partial r}                    \label{3.93} \\
e_2 &=& \frac{1}{\sqrt{1-(\alpha GM/r_0)}}\left(\frac{L}{r_0}\,
\frac{\partial}{\partial t} + \frac{E}{r_0}\,
\frac{\partial}{\partial\phi}\right)           \label{3.94} \\
e_3 &=& -\frac{1}{r_0}\frac{\partial}{\partial\theta}, 
                                               \label{3.95}
\end{eqnarray}
from which we can read off the components, 
\begin{eqnarray}
& e_0^t = \frac{E}{(1-(\alpha GM/r_0))} \hspace*{3cm} 
e_0^\phi = \frac{L}{r_0^2} &                   \label{3.96} \\
& e_1^r = \frac{1}{\sqrt{1+(\beta GM/r_0)}} & 
                                               \label{3.97} \\
& e_2^t = \frac{L}{r_0\sqrt{1-(\alpha GM/r_0)}} \hspace*{2cm} 
e_2^\phi = \frac{E}{r_0\sqrt{1-(\alpha GM/r_0)}} &
                                               \label{3.98} \\
& e_3^\theta = -\frac{1}{r_0} &                \label{3.99}
\end{eqnarray}
With the above result, we are now ready to calculate 
the Riemann curvature tensor in the Fermi normal 
coordinate system on the circular orbit. Substituting 
$f(r)=(1-\alpha\Phi(r))$ and $g(r)=(1+\beta\Phi(r))$ 
with $\Phi(r)=GM/r$ into Eqs. (\ref{3.68})-(\ref{3.73}) 
and setting $\theta = \pi/2$, the non-zero components 
of the Riemann curvature tensor in the spherical 
coordinate basis on the circular orbit of radius 
$r_0$ take the form
\begin{eqnarray}
\left. R_{trtr}\right|_{\scriptsize\mbox{orbit}} &=& 
\frac{1}{2}\alpha\Phi^{\prime\prime}(r_0) - 
\frac{\alpha\beta\left(\Phi^\prime (r_0)\right)^2}
{4(1+\beta\Phi(r_0))} + \frac{\alpha^2\left(
\Phi^\prime(r_0)\right)^2}{4(1-\alpha\Phi(r_0))} \nonumber \\
&=& \frac{\alpha GM}{r_0^3} - \frac{\alpha\beta (GM/r_0^2)^2}
{4(1+(\beta GM/r_0))} + \frac{\alpha^2(GM/r_0^2)^2}
{4(1-(\alpha GM/r_0))}  \nonumber \\
&&                                                \label{3.100} \\
\left. R_{t\theta t\theta}\right|_{\scriptsize\mbox{orbit}} 
= \left. R_{t\phi t\phi}\right|_{\scriptsize\mbox{orbit}} &=& 
\frac{\alpha r_0\Phi^\prime (r_0)}{2(1+\beta\Phi(r_0))} 
\nonumber \\
&=& -\,\frac{\alpha GM/r_0}{2(1+(\beta GM/r_0))}  \label{3.101} \\
\left. R_{r\theta r\theta}\right|_{\scriptsize\mbox{orbit}} 
= \left. R_{r\phi r\phi}\right|_{\scriptsize\mbox{orbit}} &=& 
-\,\frac{\beta r_0\Phi^\prime (r_0)}{2(1+\beta\Phi(r_0))} 
\nonumber \\
&=& \frac{\beta GM/r_0}{2(1+(\beta GM/r_0))}     \label{3.102} \\
\left. R_{\theta\phi\theta\phi}\right|_{\scriptsize\mbox{orbit}} 
&=& -\,\frac{\beta r_0^2\Phi(r_0)}{(1+\beta\Phi(r_0))} 
\nonumber \\
&=& -\,\frac{\beta GMr_0}{(1+(\beta GM/r_0))},   \label{3.103}
\end{eqnarray}
and the other components are obtained from the above 
result by index permutations. Using the components of 
the vierbein in Eqs. (\ref{3.96})--(\ref{3.99}), we 
can calculate the corresponding Riemann curvature 
tensor in the Fermi normal coordinate system from
\begin{equation}
\left. R_{abcd}\right|_{\scriptsize\mbox{orbit}} = 
e_a^\mu e_b^\nu e_c^\rho e_d^\sigma 
\left. R_{\mu\nu\rho\sigma}\right|_{\scriptsize\mbox{orbit}}, 
                                                 \label{3.104}
\end{equation}
with $R_{\mu\nu\rho\sigma}|_{\scriptsize\mbox{orbit}}$ in 
Eqs. (\ref{3.100})--(\ref{3.103}). Since our main purpose 
is to calculate the gravitoelectric and gravitomagnetic 
fields in Eqs. (\ref{3.62})--(\ref{3.63}), we will 
calculate only the components $R_{0i0j}|_{\scriptsize\mbox{orbit}}$ 
and $R_{0ijk}|_{\scriptsize\mbox{orbit}}$. 
The non-zero components are found to be
\begin{eqnarray}
\left. R_{0101}\right|_{\scriptsize\mbox{orbit}} 
&=& \frac{E^2}{(1-\alpha\Phi(r_0))^2(1+\beta\Phi(r_0))}
\left[ \frac{1}{2}\alpha\Phi^{\prime\prime}(r_0) - 
\frac{\alpha\beta\left(\Phi^\prime (r_0)\right)^2}
{4(1+\beta\Phi(r_0))} \right.  \nonumber \\
&& \left. + \frac{\alpha^2\left(
\Phi^\prime(r_0)\right)^2}{4(1-\alpha\Phi(r_0))} \right] 
- \frac{L^2}{r_0^4(1+\beta\Phi(r_0))^2}
\left[\frac{1}{2}\beta r_0\Phi^\prime(r_0)\right]   \label{3.105} \\
\left. R_{0202}\right|_{\scriptsize\mbox{orbit}} 
&=& \frac{\alpha\Phi^\prime(r_0)}{2r_0
(1-\alpha\Phi(r_0))(1+\beta\Phi(r_0))}              \label{3.106} \\
\left. R_{0303}\right|_{\scriptsize\mbox{orbit}} 
&=& \frac{E^2}{r_0^2(1-\alpha\Phi(r_0))^2
(1+\beta\Phi(r_0))}\left[\frac{1}{2}\alpha r_0
\Phi^\prime(r_0)\right] 
\nonumber \\
&& - \frac{L^2}{r_0^4(1+\beta\Phi(r_0))}
\left[\beta\Phi(r_0)\right]                         \label{3.107}
\end{eqnarray}
for $R_{0i0j}|_{\scriptsize\mbox{orbit}}$ and
\begin{eqnarray}
\left. R_{0112}\right|_{\scriptsize\mbox{orbit}} 
&=& \frac{EL}{r_0\sqrt{1-\alpha\Phi(r_0)}
(1+\beta\Phi(r_0))}\left[\frac{1}{(1-\alpha\Phi(r_0))}
\left( -\frac{1}{2}\alpha\Phi^{\prime\prime}(r_0) \right.\right. 
\nonumber \\
&& 
\left.\left. + 
\frac{\alpha\beta\left(\Phi^\prime (r_0)\right)^2}
{4(1+\beta\Phi(r_0))} - \frac{\alpha^2\left(
\Phi^\prime(r_0)\right)^2}{4(1-\alpha\Phi(r_0))} \right) 
+ \frac{\beta\Phi^\prime(r_0)}{2r_0(1+\beta\Phi(r_0))}
\right]                                             \label{3.108} \\
\left. R_{0323}\right|_{\scriptsize\mbox{orbit}} 
&=& \frac{EL}{r_0^3\sqrt{1-\alpha\Phi(r_0)}
(1+\beta\Phi(r_0))}\left[ \frac{\alpha r_0\Phi^\prime(r_0)}
{2(1-\alpha\Phi(r_0))} - \beta\Phi(r_0) \right]     \label{3.109}
\end{eqnarray}
together with other components obtained by index 
permutations for $R_{0ijk}|_{\scriptsize\mbox{orbit}}$. 
Since we are using the linearized formalism, the 
above result is valid only up to the first order in 
the metric perturbation. Thus, keeping terms up 
to the linear order in $\Phi(r_0)$ in Eqs. 
(\ref{3.105})--(\ref{3.109}), we finally obtain
\begin{eqnarray}
\left. R_{0101}\right|_{\scriptsize\mbox{orbit}} 
&=& \frac{1}{2}\alpha E^2\Phi^{\prime\prime}
(r_0) - \frac{1}{2}\beta\left(\frac{L}{r_0}\right)^2
\frac{\Phi^\prime(r_0)}{r_0}   \nonumber \\
&=& \frac{a}{L^2}\left(\frac{L}{r_0}\right)^3
\left[ E^2 + \frac{1}{2}\left(\frac{\beta}{\alpha}\right)
\left(\frac{L}{r_0}\right)^2 \right]               \label{3.110} \\
\left. R_{0202}\right|_{\scriptsize\mbox{orbit}} 
&=& \frac{1}{2}\alpha\frac{\Phi^\prime(r_0)}{r_0} 
\nonumber \\
&=& -\frac{1}{2}\frac{a}{L^2}\left(\frac{L}{r_0}\right)^3 
                                                   \label{3.111} \\
\left. R_{0303}\right|_{\scriptsize\mbox{orbit}} 
&=& \frac{1}{2}\alpha E^2\frac{\Phi^\prime(r_0)}{r_0} 
- \beta\left(\frac{L}{r_0}\right)^2\frac{\Phi(r_0)}{r_0^2} 
\nonumber \\
&=& -\frac{a}{L^2}\left(\frac{L}{r_0}\right)^3
\left[ \frac{1}{2}E^2 + \left(\frac{\beta}{\alpha}\right)
\left(\frac{L}{r_0}\right)^2 \right]              \label{3.112} \\
\left. R_{0112}\right|_{\scriptsize\mbox{orbit}} 
&=& E\left(\frac{L}{r_0}\right)\left[
-\frac{1}{2}\alpha\Phi^{\prime\prime}(r_0) + 
\frac{1}{2}\beta\frac{\Phi^\prime(r_0)}{r_0} \right] 
\nonumber \\
&=& -\frac{a}{L^2}\left(\frac{L}{r_0}\right)^4
\left[ 1+\frac{1}{2}\left(\frac{\beta}{\alpha}\right) \right]E
                                                 \label{3.113} \\
\left. R_{0323}\right|_{\scriptsize\mbox{orbit}} 
&=& E\left(\frac{L}{r_0}\right)\left[ \frac{1}{2}\alpha
\frac{\Phi^\prime(r_0)}{r_0} - \beta\frac{\Phi(r_0)}{r_0^2} \right]
\nonumber \\
&=& -\frac{a}{L^2}\left(\frac{L}{r_0}\right)^4 
\left[ \frac{1}{2} - \left(\frac{\beta}{\alpha}\right) \right]E 
                                                 \label{3.114}
\end{eqnarray}
where the expressions for $a$, $r_0$ and $E$ are 
in Eqs. (\ref{3.81})--(\ref{3.83}), and $\alpha$ and 
$\beta$ are defined in Eqs. (\ref{3.66})--(\ref{3.67}). 
Using the above result, we can calculate the gravitoelectric 
field, $(E_G)_i = R_{0i0j}|_{\scriptsize\mbox{orbit}}X^j$,
\begin{eqnarray}
(E_G)_1 = \left. R_{0101}\right|_{\scriptsize\mbox{orbit}}X^1  
                                                 \label{3.115} \\
(E_G)_2 = \left. R_{0202}\right|_{\scriptsize\mbox{orbit}}X^2  
                                                 \label{3.116} \\
(E_G)_3 = \left. R_{0303}\right|_{\scriptsize\mbox{orbit}}X^3 
                                                 \label{3.117}
\end{eqnarray}
and the gravitomagnetic field, $(B_G)_i = -\frac{1}{2}
\epsilon_{ijk}R_{jk0l}|_{\scriptsize\mbox{orbit}}X^l$,
\begin{eqnarray}
(B_G)_1 &=& -\left. R_{2303}\right|_{\scriptsize\mbox{orbit}}X^3 
                                                 \label{3.118} \\
(B_G)_2 &=& 0                                    \label{3.119} \\
(B_G)_3 &=& -\left. R_{1201}\right|_{\scriptsize\mbox{orbit}}X^1 
                                                 \label{3.120} 
\end{eqnarray}
where $X^i$ is the $i$-component of the position 
vector of the particle on which the gravitoelectric 
and gravitomagnetic forces act. It is interesting 
to observe that the above gravitomagnetic field 
is perpendicular to the moving direction of the 
observer, which is along the $x^2\equiv\phi$ 
direction, but we still do not know if this result 
holds in general. 

To get in touch with the experimental measurements, 
we have to restore the speed of light $c$ to the 
above result. We first restore $c$ to $t$ and $\tau$, 
which replaces them by $ct$ and $c\tau$. This results 
in the change of the equation of motion (\ref{3.64}) 
to
\begin{equation}
\frac{m}{c^2}\frac{d^2\vec{X}}{d\tau^2} = q_E\vec{E}_G 
+ \frac{q_B}{c}(\vec{V}\times \vec{B}_G),         \label{3.121}
\end{equation}
and the definition of $L$ in Eq. (\ref{3.75}) changes 
to
\begin{equation}
L = \frac{r_0^2}{c}\frac{d\phi}{d\tau},          \label{3.122}
\end{equation}
which means that $L$ is the angular momentum of a unit 
mass moving on a circular orbit of radius $r_0$ divided 
by $c$. We next restore $c$ to the metric tensor, which 
changes $GM$ to $GM/c^2$. Thus, the parameters in 
Eqs. (\ref{3.110})--(\ref{3.114}) take the form
\begin{eqnarray}
a &=& \frac{\alpha GM}{Lc^2}                     \label{3.123} \\
\frac{L}{r_0} &=& \frac{a}{1+\sqrt{1-3a^2}}      \label{3.124} \\
E &=& \sqrt{\left(1-a\left(\frac{L}{r_0}\right)\right)
\left(1+\left(\frac{L}{r_0}\right)^2\right)}     \label{3.125}
\end{eqnarray}
with
\begin{eqnarray}
\alpha &\equiv & 2\left(\frac{1-6\lambda}{1-4\lambda}\right) 
                                                 \label{3.126} \\
\beta &\equiv & 2\left(\frac{1-2\lambda}{1-4\lambda}\right), 
                                                 \label{3.127}
\end{eqnarray}
We see that $L$ has the dimension of length, $a$ and $E$ 
are dimensionless, and $R_{abcd}|_{\scriptsize\mbox{orbit}}$ 
has the dimension of $\mbox{length}^{-2}$. Since $X^i$ has 
the dimension of length, the dimension of the gravitoelectric 
and gravitomagnetic fields is $\mbox{length}^{-1}$, and so 
the dimensions of both sides of Eq. (\ref{3.121}) are 
consistent. 

To determine the order of magnitude of the forces in 
Eq. (\ref{3.121}), we first restore $c$ to Eq. (\ref{3.61}), 
which gives 
\begin{equation}
\frac{d^2X^i}{d\tau^2} = c^2R_{0i0j}(\tau)X^j + 
2c^2R_{ijk0}(\tau)X^k\left(\frac{V^j}{c}\right), \label{3.128}
\end{equation}
so that the gravitoelectric force per unit mass is 
\begin{equation}
F_E^i = c^2R_{0i0j}(\tau)X^j,                    \label{3.129}
\end{equation}
and the gravitomagnetic force per unit mass is 
\begin{equation}
F_M^i = 2c^2R_{ijk0}(\tau)X^k\left(\frac{V^j}{c}\right). 
                                                 \label{3.130}
\end{equation}
Since we are dealing with the situation in which things 
are almost non-relativistic, let us simplify our argument 
by considering the non-relativistic limit in which the 
Newtonian gravity holds. As mentioned before, if $v$ is 
the orbital speed of the free-falling observer in the 
circular orbit, then the orbital radius $r_0$ can be 
determined from 
\begin{equation}
\frac{GM}{r_0^2} = \frac{v^2}{r_0},              \label{3.131}
\end{equation}
which leads to $GM/r_0 = v^2$. Now, Eq. (\ref{3.122}) 
implies that 
\begin{eqnarray}
L &=& \frac{r_0}{c}\left(r_0\frac{d\phi}{d\tau}\right) 
\nonumber \\ 
&=& r_0\left(\frac{v}{c}\right).                 \label{3.132}
\end{eqnarray}
Eqs. (\ref{3.131}) and (\ref{3.132}) enable us to 
evaluate the parameter $a$ in Eq. (\ref{3.123}) as 
\begin{eqnarray}
a &\equiv & \frac{\alpha GM}{Lc^2}   \nonumber \\ 
&=& \alpha\left(\frac{v}{c}\right)               \label{3.133}
\end{eqnarray}
which is extremely small. This means that $L/r_0 
\approx a/\alpha$, and that 
\begin{equation}
E \approx \sqrt{(1-(a^2/\alpha))(1+(a^2/\alpha^2))} 
= 1 + {\cal O}(v^2/c^2) 
\end{equation}
(see Eq. (\ref{3.125})). Using these results in Eqs. 
(\ref{3.110})--(\ref{3.114}), we see that 
\begin{equation}
R_{0101},R_{0202},R_{0303} \sim \frac{a}{L^2}
\left(\frac{L}{r_0}\right)^3 = \frac{a^2}{r_0^2} = 
\frac{\alpha^2}{r_0^2}\left(\frac{v}{c}\right)^2 
                                                 \label{3.134}
\end{equation}
and 
\begin{equation}
R_{0112},R_{0323} \sim \frac{a}{L^2}
\left(\frac{L}{r_0}\right)^4 = \frac{a^3}{r_0^2} = 
\frac{\alpha^3}{r_0^2}\left(\frac{v}{c}\right)^3. 
                                                \label{3.135}
\end{equation}
Thus, Eq. (\ref{3.134}) implies that the gravitoelectric 
force per unit mass $F_E^i = c^2R_{0i0i}(\tau)X^i$ in 
our case is of the order 
\begin{eqnarray}
F_E^i &\sim & c^2\frac{\alpha^2}{r_0^2}
\left(\frac{v}{c}\right)^2 X    \nonumber \\
&=& \alpha^2 \left(\frac{v}{r_0}\right)^2 X  \nonumber \\
&=& \frac{4\pi^2\alpha^2 X}{T^2},               \label{3.136}
\end{eqnarray}
where $X$ is the distance between the particle and 
the observer, and $T$ is the period of rotation of 
the observer, satisfying $v = 2\pi r_0/T$. On the 
other hand, Eqs. (\ref{3.130}) and (\ref{3.135}) 
imply that the gravitomagnetic force per unit mass 
is of the order 
\begin{eqnarray}
F_M^i &\sim & c^2\frac{\alpha^3}{r_0^2}
\left(\frac{v}{c}\right)^3 \left(\frac{V}{c}\right) X
\nonumber \\
&=& \alpha \left(\frac{vV}{c^2}\right) 
\frac{4\pi^2\alpha^2 X}{T^2},                   \label{3.137}
\end{eqnarray}
where $V$ is the particle's speed relative to the 
observer. We see that the gravitomagnetic force is 
very much smaller than the gravitoelectric force by 
a factor of $\alpha vV/c^2$. 

To appreciate how small these forces are, let us 
consider an observer sitting inside a satellite 
moving in the geostationary orbit. Such a satellite 
has a speed $v$ of about 3 km/s and an orbital period 
$T$ of about 24 hours (see page 21 of \cite{Satellite}). 
By using these numbers in Eqs. (\ref{3.136}) and 
(\ref{3.137}) and assuming that the distance between 
the particle and the observer is about 1 m, it is 
easy to show that the magnitude of the gravitoelectric 
force is of the order $10^{-8}$ N. Since $v/c\approx 
10^{-5}$, the size of the gravitomagnetic force is 
smaller than that of the gravitoelectric force by 
at least 13 orders of magnitude if we assume that 
the particle speed is about 1 m/s. Despite the 
smallness of these forces, the phenomenon of 
gravitoelectromagnetism in General Relativity 
has been experimentally tested and confirmed in 
the Gravity Probe B experiment \cite{GravityProbeB}. 
From Eqs. (\ref{3.136})--(\ref{3.137}), it can be 
seen that the parameter $\lambda$ in Rastall gravity 
enters the expressions of the forces through the 
parameter $\alpha = 2(1-2\lambda + {\cal O}(\lambda^2))$. 
This causes a shift in the magnitudes of the 
gravitoelectric and gravitomagnetic forces by 
about some number times $100\lambda$ percent. 
Thus, by investigating the experimental errors 
in the Gravity Probe B experiment and by performing 
a more realistic calculation of the 
gravitoelectromagnetism, it is expected that one 
should be able to estimate the upper bound of 
$\lambda$. This concludes our discussion of the 
gravitoelectromagnetism in the linearized Rastall 
gravity.




\chapter{Conclusions}

In this report, we have studied the linearized 
Rastall gravity and one of its phenomenological 
consequences, known as the gravitoelectromagnetism. 
Beginning with the application of the linearized 
formalism in General Relativity in the analysis of 
the weak gravitational field in Rastall gravity in 
Chapter 2, it was found that, by using a suitable 
choice of coordinate system, the Rastall-gravity 
equation reduces to the wave equation satisfied 
by the (modified) metric perturbation with the 
energy-momentum-stress tensor playing the role 
of a source term, just like what is typical in 
General Relativity. Once the general form of the 
solution to the wave equation has been obtained 
by using the method of Green's function, we used 
it to calculate the metric tensor for the matter 
without pressure and shear stress. The metric 
tensor that we obtained depends on the parameter 
$\lambda$, and reduces to the one in General 
Relativity in the limit $\lambda\to 0$. 

In Chapter 3, we used the solution obtained in 
Chapter 2 to analyze the phenomenon of 
gravitoelectromagnetism as seen by an observer 
moving along a circular orbit in a spherically 
symmetric gravitational field. In order to prepare 
for such an analysis, we presented a review of 
a mathematical machinery, known as the Fermi normal 
coordinate system, which is the coordinate system 
that a free-falling observer uses to describe 
things nearby him. It was found that this 
free-falling observer will see ``another free 
particle" as being subject to both 
velocity-independent and velocity-dependent 
forces. To the lowest order in the particle 
velocity divided by the speed of light, these 
forces mimic the electric and magnetic forces 
in electromagnetism, and therefore, are called 
the gravitoelectric and gravitomagnetic forces. 
From our result, we observed that an observer 
moving along a circular orbit will measure the 
gravitomagnetic field as being perpendicular 
to his moving direction, but we do not know 
if this result is true in general. We also 
discussed the orders of magnitude of these 
forces, and mentioned the possibility of 
obtaining the upper bound of $\lambda$ from 
the experimental data. 

We end this report with a remark that our analysis 
of the gravitoelectromagnetism was limited to 
the observer moving along a circular orbit. This 
is due to the fact that the circular orbit is the 
only orbit whose form we know exactly. As the other 
kinds of orbits are generally not closed, the only 
way to analyze the corresponding gravitoelectric 
and gravitomagnetic fields is to use the computer 
to generate the observer's path and the vierbein 
along it step by step, and subsequently obtain 
the gravitoelectromagnetic fields. This is beyond 
the scope of our work, and is left as a future 
work.




\end{document}